\documentclass[sn-mathphys]{sn-jnl}% Math and Physical Sciences
\usepackage{bm}
\usepackage{amssymb}
\usepackage[utf8]{inputenc}
\usepackage{graphicx}
\usepackage{hyperref}
\usepackage{float}
\usepackage{amsmath}
\usepackage[T1]{fontenc}
\usepackage{hyperref} %% for \autoref
\usepackage{isotope} %% for \isotope
\usepackage{upgreek}
\usepackage{sidecap}
\usepackage{wrapfig}
\usepackage{placeins}
\usepackage{booktabs}
\usepackage{color}
\usepackage{mathrsfs}
\usepackage{epsfig}
\usepackage{bm}
\usepackage{ulem,color,graphics}
\usepackage{multirow}
\usepackage{lineno}
\usepackage{url}
\def\bit{\begin{itemize}}
\def\eit{\end{itemize}}
\def\bnu{\begin{enumerate}}
\def\enu{\end{enumerate}}
\def\sss{\scriptscriptstyle}

\def\F {{{\cal F}}}

\def\O {{{\cal O}}}

\def\nn{\nonumber }

\def\M {{{\cal M}}}
\def\x{\times}
\def\Ket#1{\vert\vert#1 \rangle}
\def\Bra#1{\langle #1\vert\vert}

\def\ie{{\textit i.e., }}

\def\nn{\nonumber }
\def\be{\begin{equation}}
\def\ee{\end{equation}}
\def\br{\begin{eqnarray}}
\def\er{\end{eqnarray}}
\def\brn{\begin{eqnarray*}}
\def\ern{\end{eqnarray*}}

\def\jb{ {\bf j}}

\def\ket#1{\vert#1 \rangle}
\def\rf#1{{(\ref{#1})}}

\def\sixj#1#2#3#4#5#6{\left\{\negthinspace\begin{array}{ccc}
#1&#2&#3\\#4&#5&#6\end{array}\right\}}
\def\ninj#1#2#3#4#5#6#7#8#9{\left\{\negthinspace\begin{array}{ccc}
#1&#2&#3\\#4&#5&#6\\#7&#8&#9\end{array}\right\}}

\def\go{\rightarrow  }
\def\etal {{\textit et al.}}
\def\E {{{\cal E}}}

\def\J {{{\cal J}}}
\def\X {{{\cal X}}}

\def\b {{\beta}}

\def\sss{\scriptscriptstyle}

\def\Lh {\hat{L}}

\def\E {{{\cal E}}}
\def\w {{\omega}}
\def\etal {{\it et al.}}
\def\b {{\beta}}
\def\d {{\dagger}}%
\jyear{2023}%
\begin{document}
\title{
 %Neutrinoless $\beta\beta$-Decay in 
%Double Charge Exchange  Quasiparticle Tamm-Dancoff Approximation 
%A New Nuclear Model for  Neutrinoless $\beta\beta$-Decay 
Neutrinoless $\beta\beta$-Decay in DCEQTDA
}
%\ and  Double Charge-Exchange Resonances}
%
\author[1]{\fnm{C. De Conti} }

\author[2,3]{\fnm{V. dos S. Ferreira} }

\author[2]{\fnm{A.R. Samana} }

\author[4,5]{\fnm{C.A. Barbero} }

\author[4]{\fnm{F. Krmpoti\'c} }

\affil[1]{S\~ao Paulo State University (UNESP), School of Engineering and Sciences, 19274-000 Rosana, SP, Brazil}

\affil[2]{Departamento de Ci\^encias Exatas,
Universidade Estadual de Santa Cruz, Campus Soane Nazar\'e de Andrade,
Rod. Jorge Amado Km 16, 45662-900 Ilh\'eus, BA, Brazil}

\affil[3]{Centro de Ci\^encias Exatas e Tecnol\'ogicas, Universidade 
	Federal do Rec\^oncavo da Bahia, 44380-000 Cruz das Almas, BA,  Brazil}

\affil[4]{Departamento de F\'isica, Universidad Nacional de La Plata,
C.C. 67, 1900 La Plata, Argentina}

\affil[5]{Instituto de F\'isica La Plata, CONICET, 1900 La Plata, Argentina}

%%%%%%%%%%%%
\abstract{
We have recently developed a nuclear model, which is a natural extension of 
the $pn$-QRPA model, specially designed to describe double charge exchange 
(DCE) processes generated by two-body DCE transition operators. It is based 
on the Quasiparticle Tamm-Dancoff Approximation (QTDA) for $pn$ and $2p2n$ 
excitations in intermediate and final nuclei, respectively, and will be called DCEQTDA.

As such, this model, having the same number of free parameters as the $pn$-QRPA, 
also brings into play the excitations of four quasiparticles to build up the final nuclear 
states, which are then used to evaluate the nuclear matrix elements (NMEs) for 
all  $0^+$ and $2^+$ final states, including resonances, and not just for the 
ground state as in  $pn$-QRPA.
In addition, it allows us to evaluate: (a) the  values of $Q_{\b\b}$, 
(b) the excitation energies in final nuclei, and (c) the DCE sum rules, 
which are fulfilled in the DCEQTDA.
So far, this model has been used mainly to calculate double beta decays with 
the emission of two neutrinos ($2\nu\beta\beta$-decay). Here,  we extend it to 
the study of these processes when no neutrinos are emitted  ($0\nu\beta\beta$-decay), 
evaluating them in a series of nuclei, but paying special attention to
 (i) $^{76}$Se, which have been measured recently in the GERDA  
 and MAJORANA experiments, and (ii) $^{124}$Te, for which the first direct 
 observation of the double electron capture $2\nu$ has been performed with 
 the XENON1T dark matter detector. We obtain good agreement with the data 
 for both the ground state and the excited states. 
  The validity of the DCEQTDA model is checked by comparing  
  the calculation with the experimental data for the $2\nu\b\b$ NMEs, and for 
  the $Q_{\b\b}$, in a series of nuclei.
  }

\noindent

%\pacs{14.60.Lm,21.60.-n, 23.40.Bw, 23.40.-s}


\maketitle

\newpage

\section{ Introduction}
\label{Sec1}

 If the driving mechanism of the $0\nu\b\b$-decay is through the exchange 
 of a left-handed light neutrino of Majorana type, which is forbidden in the 
 Standard Model, its detection will allow us to find the effective mass of the 
 neutrino $\langle m_ {\nu}\rangle$, provided we know the nuclear matrix 
 element (NME) $M^{0\nu}$, since the decay {\color{black}  
amplitude} {\color{black} rate}   is proportional to 
$\vert \langle m_{\nu} \rangle M^{0\nu} \vert^2 $.  
%$ |\langle m_{\nu} \rangle M^{0\nu}|^2 $.  
%arturo

These NMEs are calculated using different nuclear models, and there are no experimental data or model-independent sum rules to corroborate their calculated values.
As a consequence, a large scatter in the results (up to factors of three) has emerged.

Strictly speaking, despite the enormous efforts invested in these calculations, we are still not entirely sure of the order of magnitude of the  NMEs  $M^{0\nu}$\cite{Bra21}.
We also do not know what is the total transition intensity of the $0\nu\b\b$-decay, and what is the proportion of it that ends into the ground state. We will try to give an answer to these questions. We know this only for the double Gamow-Teller operator from the work of Auerbach \etal  \cite{Zhe89,Aue89,Aue18,Aue18a}, but not for $M^{0\nu}$.

%In recent years much attention is being paid to $\b\b$-decays  to excited states in final nuclei, and there are several large underground experiments operating for detection of $\langle m_{\nu}\rangle$, such as the GERDA  and MAJORANA searches in $^{76}$Ge~\cite{Agos20,Arn21}.

In recent years much attention has been paid to  $\b\b$-decays to excited states in final nuclei, operating several large underground experiments, such as the GERDA \cite{Agos20}, MAJORANA \cite{Arn21}, and LEGEND \cite{LEGEND} searches in $^{76}$Ge, the first two of which have already been completed.

On the other hand,
 the NUMEN heavy ion multidetector, designed for a complementary approach to $0\nu\b\b$   NMEs,
 is currently taking data \cite{Len19,Cav20,Fin20,Ago20,Cap20}.
Theoretical comparisons have been made \cite{Shi18,Men18,San18,San20} between the $0\nu\b\b$ NME and the double Gamow-Teller transition to the ground state of the final nucleus, where a very small portion of the total strength is found.

In addition to evaluating  the $\b\b^-$-decays of $^{76}$Ge, we also discuss the $\b\b^+$-decays of $^{124}$Xe, where has been  done the first direct observation of the $2\nu$ double electron capture  ($2\nu ee$) for the ground state  in 
$^{124}$Te, with the half-life ${\tau_{2\nu}^{ee}}(0^+_1) = (1.8\pm 0.6)\times 10^{22}$ y \cite{Apr19}, which is the longest half-life ever measured directly, about one trillion times the age of the Universe\footnote{As pointed out by Doi~\cite{Doi93} in order to satisfy the energy-momentum conservation this decay mode should be accompanied by an emission of some additional particle(s).}.
More recently,  by a XMASS experiment \cite{Hir20} has been imposed  the constraint on this half-life ${\tau_{2\nu}^{ee}}(0^+_1) > 2.1 \times 10^{22}$ y at $90\%$  confidence level, which is consistent with above measurement.

%{\color{magenta}  ------------------------------------------
\section {General Formalism for DCE Processes }
\label{Sec2}

\subsection {Single and Double Charge Exchange Strengths and their Sum Rules  }\label{Sec2A}

The  Single Charge Exchange (SCE) operators $\O_0^\mp=\tau^\mp$, and $\O_1^\mp=\tau^\mp\sigma$,
play  a fundamental role in $\b^\mp$-decays: $(A,Z)\rightarrow (A,Z\pm 1)$. In the same way, 
the Double Charge Exchange (DCE) operators  $ \mathscr{  D}_{J\J}^\mp=[\O^\mp_{J}\x\O^\mp_{J}]_{\J}$,
with $J=0,1$, and   $\J= 0,2$, are essential for 
the $\b\b^\mp$-decays: $(A,Z)\rightarrow (A,Z\pm 2)$, and  closely related to the  $0\nu$ and  $2\nu$  NMEs.
In turn, the operators $\O_J$, and   $ \mathscr{  D}_{J\J}^\mp$ are related to the
corresponding SCE, and DCE  reaction processes, and their giant resonances.

%, since they can be observed experimentally only through nuclear reactions. 
%The summation  goes over all intermediate virtual states
%$\ket{ J_i}$ in both  nuclei $(A,Z\mp 1)$. 

An attempt is made to infer the value of $M^{0\nu^-}(0^+_1)$ from heavy-ion reaction data at low momentum transfer using 
$M _{10}$ and $M _{00}$ for $M^{0\nu}_{A}$ and $M^{0\nu}_{V}$, respectively \footnote{(See for example \cite[Eq. (16)]{Cap15}).},
although these two sets of NMEs differ from each other in several respects, as will be discussed below.

The one-body matrix elements matrix elements are usually expressed as 
\br
\Bra{J^+_i}\O^\pm_J\Ket{0^{+}}=\sum_{pn}\rho^\pm(pnJ_i^\pi) W_{J0J}(pn),
\label{1} \er
and
\br
\Bra{\J^+_f}\O^\pm_J\Ket{J^+_i}=\sum_{np}\rho^\pm(pnJ_i,\J^+_f) W_{J0J}(pn),
\label{2} \er
\ie as a product of model dependent one-body densities
\br
\rho^{-}(pnJ_i)&=&
\hat{J}^{-1}\Bra{{J_i}}(c^{{\dagger}}_{p}c_{\bar{n}})_{J}\Ket{0^+_ i},
\nn\\
\rho^{+}(pnJ _i) &=&
\hat{J}^{-1} \Bra{J_i}(c^{{\dagger}}_{n}c_{\bar{p}})_{J}\Ket{0^+_i},
\label{3} \er
from the initial to intermediary states, and
\br
\rho^{+}(pnJ _i,\J_f)&=&
\hat{J}^{-1} \Bra{\J_f}(c^{{\dagger}}_{n}c_{\bar{p}})_{J}\Ket{{J_i}},
\nn\\
\rho^{-}(pnJ _i,\J_f)&=&
\hat{J}^{-1} \Bra{\J_f}(c^{{\dagger}}_{p}c_{\bar{n}})_{J}\Ket{{J_i}},
\label{4} \er
from the intermediary to final states,
and 
of the single-particle NMEs
\footnote{We use here the angular
momentum coupling scheme
$\ket{({1 \over 2},l)j}$\cite{Krm94}.}:
\begin{eqnarray}
&&W_{LSJ}(pn)\equiv\sqrt{4 \pi} \Bra{{\rm p}} O_{LSJ}\Ket{{\rm n}}
\nn\\
&=&\sqrt{2}\hat{S} \hat{J}\hat{L}\hat{l}_n\hat{j}_n\hat{j}_p
(l_nL | l_p) 
\ninj{l_p}{{1 \over 2}}{j_p}{L}{S}{J}{l_n}{{1 \over 2}}{j_n},
%\nn\\
\label{5}
\end{eqnarray}%15
of the  non-relativistic 
$\b$-decay operators
\br
 O_{LSJ}= (Y_{L} \otimes\sigma^{S})_{J}, 
\label{6}\er
with $S=0$ (1) for vector (axial vector) operators. Note that  $ O_{0JJ}\equiv O_J$, and $(l_nL|l_p)\equiv (l_n0L0|l_p0)$.

%Thus %, the $2\nu\b\b$ NMEs are expressed in the form

The nuclear models are employed  
to evaluate the total SCE strengths 
\br
S^{\{\mp 1\}}_{J}&\equiv&\sum_{i}
|\Bra{J_i}\O^\mp_{J}\Ket{0^+}|^2\equiv\sum_{i}
{ {\textsf s}^{\{\mp 1\}}_{J}(J_i)},
\label{7}\end{eqnarray}
going from the initial ground state $\ket{0^+}$ 
to  intermediate states  $\ket{J_i^+}$, as well as the total DCE strengths 
\br
S^{\{\mp 2\}}_{J\J}&=&
\sum_{ f}|\Bra{\J_f}\mathscr{D}_{J\J}^\mp\Ket{0^+}|^2\equiv\sum_{ f} {\textsf s} ^{\{\mp 2\}}_{J\J}(\J_f),
\label{8}
\end{eqnarray}
going from  $\ket{0^+}$ to final states $\ket{\J_f}$.

When both $\ket{J_i}$ and $\ket{\J_f}$ are 
complete sets of  states
that can be reached by operating with $\O^\pm_J$, and   $\mathscr{D}_{J\J}^\pm$
on  $\ket{0^+}$,  the strength differences
\br
&&{ S}^{\{ 1\}}_{ J}=  
S^{{\{- 1\}}}_{ J}- S^{{\{+ 1\}}}_{ J},
\nn\\
&&{S}^{\{ 2\}}_{J\J}=  
S^{{\{-2\}}}_{J\J}- S^{{\{+ 2\}}}_{J\J},
\label{9}\end{eqnarray}
%obey the  SCE Ikeda's sum rules \cite{Ike63} 
should obey the  SCE Ikeda's sum rules \cite{Ike63}
\br
{\textsf S}^{\{ 1\}}_{ J}
&=&(2J+1)(N-Z),
\label{10}\er
and the DCE sum rules \cite{Vog88,Mut92,Zhe89}
\begin{eqnarray}
&& {\textsf S}^{\{ 2\}}_{00} =2(N-Z)(N-Z-1),
\nn\\
&&  {\textsf S}^{\{ 2\}}_{10}
=2(N-Z)\left(N-Z+1+2S^{\{-1\}}_1\right)
-\frac{2}{3}C\nn, \\
&& {\textsf S}^{\{ 2\}}_{12}
=10(N-Z)\left(N-Z-2+2S^{\{-1\}}_1\right)+\frac{5}{3}C,
\label{11}
\end{eqnarray}
where $C$ is a relatively small positive quantity, given by~\cite[Eq. (4)]{Mut92}.
Since the terms proportional to $C$ will not considered in the present work, the following inequalities arise:
\br
 {\textsf S}^{\{ 2\}}_{10}& \ge&{ S}^{\{ 2\}}_{10},\hspace{1cm} {\textsf S}^{\{ 2\}}_{12} \le{ S}^{\{ 2\}}_{12}.
\label{12}\er
It is desirable that any nuclear model used to calculate DCE processes satisfies the sum rules \rf{11} for the total strengths $S^{\{\mp 2\}}_{J\J}$. 
Through them, we can evaluate the fractions of strengths going to the individual states 
as
\br
R^{\{\mp 1\}} (J_i) &=&{{\textsf s} ^{\{\mp 1\}}_{J}(J_i)}/{S^{\{\mp 1\}}_{J}}, 
\nn\\
R^{\{\mp 2\}}_{J\J} (\J_f) &=& {{\textsf s} ^{\{\mp 2\}}_{J\J}(\J_f)}/{S^{\{\mp 2\}}_{J\J}}, 
\label{13}\end{eqnarray}
calibrating in this way  the  NMEs of  the operators  $\O_J$ and $ \mathscr{  D}_{J\J}$.
Later,  the same will be done with the $0\nu\b\b$ NMEs, which 
will allow us to determine their order of magnitude.

%{\color{blue} 
The main difference between (${ S}^{\{ 1\}}_{ J}, {S}^{\{ 2\}}_{J\J})$,  and (${\textsf S}^{\{ 1\}}_{ J}, {\textsf S}^{\{ 2\}}_{J\J})$ is that while the former depend on nuclear structure the latter do not. These two quantities in principle should be equal to each other, and it is desirable that any nuclear model used to calculate the SCE and  DCE processes
satisfies the condition.

This generally occurs in most theoretical calculations with the Ikeda's sum rule,
but not with ${\textsf S}^{\{ 2\}}_{J\J}$ because of \rf{12}.
%}
%\newpage

When the above condition on the completeness of the basis $\ket{J_i}$ is satisfied, we get\br
\Bra{\J_f}\mathscr{D}_{J\J}^\mp\Ket{0^+}
&=&\sum_{i}\Bra{\J^+_f}\O^{\pm}_J\Ket{J_i}
\Bra{J_i}\O^\pm_J\Ket{0^{+}}
\nn\\
&=&\sum_{i}\sum_{p_1n_1p_2n_2}\varrho^\mp(p_1n_1p_2n_2;J_i,\J_f)
W_{J0J}(p_1n_1) W_{J0J}(p_2n_2),
\nn\\
&\equiv&\sum_{i}\Bra{\J_f}\mathscr{D}_{J_i\J}^\mp\Ket{0^+}
\label{14}\end{eqnarray}%13
with $J=0,1$,  and $\J= 0,2$, and
 \br
\varrho^\mp(p_1n_1p_2n_2;J_i,\J^+_f)&=&\varrho^\mp(p_1n_1;J_i)\varrho^\mp(p_2n_2;J_i,\J^+_f),
%\nn\\
\label{15} \er%14

Given that the reaction matrix  elements   $\Bra{\J_f}\mathscr{D}_{J\J}^\mp\Ket{0^+}$ do not depend on the energies of the intermediate $\ket{J_i}$ states, we can
simplify their evaluation, using
\br
M^\mp_J(\J_f)\equiv\Bra{\J_f}
\mathscr{D}_{J\J}^\mp\Ket{0^+}
&=&\hat{J}^{-1}\sum_{p_1n_1p_2n_2}
\bar\varrho^\mp(p_1n_1p_2n_2;J\J^+_f)
\nn\\
&\x&
%\bar\varrho^\mp(p_1n_1p_2n_2;J\J^+_f)
W_{0JJ}(p_1n_1) W_{0JJ}(p_2n_2).
%\nn\\
\label{16}\end{eqnarray}
where
\br
&&\bar\varrho^\mp(p_1n_1p_2n_2;J\J^+_f)=\sum_i \varrho^\mp(p_1n_1p_2n_2;J_i\J^+_f )
\nn\\
&\equiv&\sum_i\varrho^\mp(p_1n_1;J_i)\varrho^\mp(p_2n_2;J_i,\J^+_f)
%\varrho^\mp(p_1n_1p_2n_2;J_i,\J^+_f)&=&\varrho^\mp(p_1n_1;J_i)\varrho^\mp(p_2n_2;J_i,\J^+_f),
\label{17}\er

\subsection {Two-neutrino Double Beta Decay  }\label{Sec2B}

The $2\nu\b\b$-decay processes can occur, if energetically allowed, as a second order perturbation of the weak Hamiltonian, independently of whether neutrinos are Dirac or Majorana and massive or massless allowed %\cite{Hax84,Doi93}. 
\cite{Doi93}. 
Their matrix elements  read
\br
 &&M^{2\nu^\mp}_{J}(\J_f)=\frac{-g_{J}^2}{\hat\J}
\sum_{i }
\frac{
\Bra{\J_f}\mathscr{D}_{J_i\J}^\mp\Ket{0^+}
}
 {\left({\cal D}_{J_i,\J_f}^{2\nu^\mp}\right)^{\J+1}},
%\nn\\
\label{18}\er %12
for  F ($g_{0}\equiv g_{\sss V}$), and GT ( $g_{1}\equiv g_{\sss A}$)
transitions. Moreover, $\hat{\J}=\sqrt{2\J+1}$, and  $\ket{\J^+_f}$ are the final states
with $\J=0$, and $\J=2$.

The energy denominator  is~
\begin{eqnarray}
{\cal D}_{J_i\J_f}^{\mp}
&=&E_{J_i }^{\{\mp 1\}}-E_{0^+}^{\{0\}}+\frac{E_{0^+}^{\{0\}}-E^{\{\mp 2\}}_{\J_f}}{2},
\nn\\
&=&E_{J_i }^{\{\mp 1\}}-\frac{E_{0^+}^{\{0\}}+E^{\{\mp 2\}}_{\J_f}}{2}
\label{19}\end{eqnarray}%16
where $E_{0^+}^{\{0\}}$, $E_{J_i}^{\{\mp 1\}}$, and $E_{\J_f}^{\{\mp 2\}}$
 are, respectively,  the energies of: 
(i) the ground state of the decaying nucleus $(A, Z)$,
 (ii) the $J_i ^{+}$ state in the intermediate nucleus $(A,Z\pm 1)$, and
 (iii) the $\J_f^{+}$ state in the final nucleus $(A,Z\pm 2)$.
The term $({E_{0^+}^{\{0\}}-E^{\{\mp 2\}}_{\J_f}})/2$ comes from the approximation done for the energy   of the  leptons $e+\nu$, emitted in the first $\b$ decay. 
It is equal to half of the corresponding $Q$-value (see \rf{39}).

\subsection {Neutrinoless  Double Beta Decay  }\label{Sec2C}

If neutrinos are assumed to be Majorana particles, then the $0\nu\b\b$-decay mode can take place under some conditions, which are explained by Doi~\cite{Doi93}.

The  $0\nu\b\b$ NMEs for the  final $0^+$ and $2^+$ states
are completely different from  each other, since they are spawned from different
parts of the vector (V), and axial-vector (A) weak-hadronic-currents
$J^{\mu}=\left(\frac{}{}\rho,{\jb}\right)$; see, for instance  \cite[Eqs. (7), (8)]{Bar98}.  
In fact, while   the first comes from $\rho_V$, and $\jb_A$, which are velocity independent parts of $J^{\mu}$ ($\rho_V$, and $\jb_A$), and 
are akin to  $\mathscr{D}_{00}$, and $\mathscr{D}_{10}$, respectively, the  second
comes from the velocity dependent parts of  $J^{\mu}$ ($\rho_A$, and $\jb_V)$, that do not
resemble $\mathscr{D}_{12}$; see, \cite[Eqs.(5), (6)]{Tom00}. 
Because of this,  $ 0\nu\b\b$-decays  to final states $2^+$ 
are not considered here.

In the case of $0^+$ final states, pseudo-scalar (P), and weak magnetism (M)  induced currents also contribute 
to the $0\nu\b\b$-decay, and the corresponding NME, $M^{0\nu^{\mp}}(0^+_f)$,  is usually 
presented as a sum of F, GT, and Tensor (T) parts (see, for example, \cite [Eq. (3)] {Sen13}). 
The term F arises only from the V weak current, but the  GT and T terms contain mixtures of 
A, M, and P currents (see Refs. \cite{Krm94,Bar98,Bar99,Fer17}).
%(see \cite[Eq. (4.5)]{Fer17}). 
We prefer to highlight the individual contributions of the weak currents, and  write
\br
M^{0\nu^{\mp}}(0^+_f)
&=&\sum_{X=V,A,M,P}M^{0\nu^{\mp}}_{X}(0^+_f),
\label{20}\er
%where $X=V,A,M,P$.  %{\color{blue} 
 The  weak  coupling constants $g_{\sss V}$,
$g_{\sss A}$,
 $f_{\sss M}=(g_{\sss M}+g_{\sss V})/(2M_N)$, and 
 $g'_{\sss P}=g_{\sss P}/(2M_N)$ are incorporated within $M^{0\nu^{\mp}}_{X}(0^+_f)$.
They  are fixed as follows:
$g_{\sss V}=1$,  and  $g_{\sss M}=3.7$ from  Conservation of Vector
Current, $g_{\sss A}=1.27$ from the experimental data~\cite{Ber12},
and $g_{\sss P}=2M_Ng_{\sss A}/(q^2+m_\pi^2)$
from the  assumption of Partially Conserved Axial Current ~\cite{Wal95}.

The Finite Nucleon Size effects are introduced
through the usual  dipole  form factors
\begin{eqnarray}
g_{\sss V}&\go&g_{\sss V}(k^2)\equiv g_{\sss V}\Lambda^4_V(\Lambda^2_V+k^2)^{-2},
\nn\\
g_{\sss A}&\go&g_{\sss A}(k^2)\equiv g_{\sss A}\Lambda^4_A(\Lambda^2_A+k^2)^{-2},
\nn\\
f_{\sss M}&\go&f_{\sss M}(k^2)\equiv f_{\sss M}\Lambda^4_V(\Lambda^2_V+k^2)^{-2},
\nn\\
g'_{\sss P}&\go&g'_{\sss P}(k^2)\equiv g'_{\sss P}\Lambda^4_A(\Lambda^2_A+k^2)^{-2},
\label{21}\end{eqnarray}%17
where $\Lambda_V=0.85$ GeV, and $\Lambda_A=1.086$ GeV are
the cut-off parameters as  found in~\cite{Krm92,Sim99,Yao15}.  
The Short Range Correlations are included in the way indicated in \cite[Eqs. (2.29)-(2.31)]{Fer17}.

We will also discuss the NMEs $M^{0\nu^{\mp}}_{V_0}$ and  $M^{0\nu^{\mp}}_{A_1}$ which are the parts of $M^{0\nu^{\mp}}_{V}$ and  $M^{0\nu^{\mp}}_{A}$  engendered, respectively, only
by the intermediate states $J_i^{\pi}=0^+_i$ and $1^+_i$.
%}

Like the NMEs $M^{2\nu^\mp}_{J}(\J_f)$,  the $M^{0\nu^\mp}_X(0^+_f)$ in \rf{20} can be expressed by the density matrix $\varrho^\mp$, as follows
\br
M^{0\nu^\mp}_X(0^+_f)
&=&\sum_{J_i}\sum_{p_1p_2n_1n_2}\varrho^\mp(p_1n_1p_2n_2;J^\pi_i0_f)
\nn\\
&\x&m^{0\nu}_{X}(p_1n_1p_2n_2;J,{\cal D}_{J_i,0_f}),
\label{22}\er
where $m^{0\nu}_{X}(p_1n_1p_2n_2;{\cal D}_{J_i,0_f})$ are  
the single particle $0\nu\b\b$ NMEs, which do not depend on nuclear models.

Because the $0\nu\b\b$  NMEs depend only very weakly on the energy 
denominators ${\cal D}_{ J^\pi_i,0^+_f}$, calculations are usually performed in the 
Closure Approximation (CA), where these are approximated by a constant 
value ${\cal D}$ of the order of $10$ MeV~\cite{Sen13}.
This greatly simplifies the numerical calculations, and it will be done here. 
Eq. \rf{22} becomes
\br
M^{0\nu^\mp}_X(0_f)
&=&\sum_{J}\sum_{p_1p_2n_1n_2}\bar\varrho^\mp(p_1n_1p_2n_2;J,0_f)
\nn\\
&\x&m^{0\nu}_{X}(p_1n_1p_2n_2;J,{\cal D}),
\label{23}\er
where the densities $\bar\varrho^\mp(p_1n_1p_2n_2;J,0_f)$ are given by \rf{17}.

The single particle $0\nu\b\b$ NMEs can be derived 
from  ~\cite [Eqs. (20)]{Fer17}, and they are:
\newpage
\br
m^{0\nu}_{V}(p_1p_2n_1n_2;J,{\cal D})
&=&W_{J0J}(p_1n_1) W_{J0J}(p_2n_2)\nn\\
&\x&
{\cal R}^V_{JJ}(p_1n_1p_2n_2{\cal D}),
\label{24}\er
%M_A}
\br
m^{0\nu}_A(p_1p_2n_1n_2;J,{\cal D})
&=&\sum_{L}W_{L1J}(p_1n_1) W_{L1J}(p_2n_2)
\nn\\
&\x&(-)^{L+1}
{\cal R}^{A}_{LL}(p_1n_1p_2n_2;{\cal D}),
%\nn\\
\label{25}\er
%M_P}
\br
m^{0\nu}_P(p_1p_2n_1n_2;J,{\cal D})
&=&-\sum_{ LL'l}(-)^{J+(L+L')/2}\Lh\Lh'(LL'll)
\nn\\
&\x&(11\vert l)  W_{L1J}(p_1n_1) W_{L'1J}(p_2n_2)
\nn\\
&\x&\sixj{L}{L'}{l}{1}{1}{J}{\cal R}^P_{LL'}(p_1n_1p_2n_2;{\cal D}),
%\nn\\
\label{26}\er
and
%M_M}
\br
m^{0\nu}_M(p_1p_2n_1n_2;J,{\cal D})
=-\sum_{ LL' l}(-)^{J+(L+L')/2}\Lh\Lh'(LL' \vert l)
&&
\nn\\
\x (11\vert l) ~W_{L1J}(p_1n_1) ~W_{L'1J}(p_2n_2)\left[2-\frac{l(l+1)}{2}\right]
&&
\nn\\
\x \sixj{L}{L'}{l}{1}{1}{J}
{\cal R}^{M}_{LL'}(p_1n_1p_2n_2;{\cal D}).
%\nn\\
\label{27}\er
We note that $m^{0\nu}_V$, and $m^{0\nu}_A$ are usually written 
as $m^{0\nu}_F$, and $m^{0\nu}_{GT}$, and that the  $l=2$ parts in $m^{0\nu}_P$, 
and $m^{0\nu}_M$ are the tensor components of  $M^{0\nu}$.

The two-body radial integrals are defined as (see~\cite{Bar98})
\begin{eqnarray}
{\cal R}_{LL'}^X(p_1n_1p_2n_2;{\cal D})&=& r_N\int dk~k^{2}
v_X(k;{\cal D}) 
\nn\\
&\x&  
R_L(p_1n_1;k) ~R_{L'}(p_2n_2;k),
%\nn\\
\label{28}\end{eqnarray}
with
\begin{eqnarray}
R_L(pn;k) &=&
\int_0^\infty u_{n_pl_p}(r)u_{n_nl_n}(r)~j_L(kr)~r^{2} ~dr,
%\nn\\
\label{29}\end{eqnarray}
 and
\begin{eqnarray}
v_X(k;{\cal D})&=&\frac{2}{\pi}\frac{{\mathscr G}_X(k)}{k(k+{\cal D)}},
\label{30}
\end{eqnarray}
where ${\mathscr G}_X(k)=g^2_V(k),  g^2_A(k),   k^2 f^2_M(k)$,  and $k^2 g_P(k)[2g_{\sss A}(k)-k^2g_P'(k)]$,
for $X=V,A,M$, and $P$, respectively. The ratio
 $r_N = 1.2A^{1/3}$ fm is introduced to make the $0\nu\b\b$ NMEs dimensionless \cite{Doi93}.

The $\b\b$-decay  half-lives are evaluated from
\br
[\tau(\J_f)]^{-1}=
 \F^2 \left\vert M(\J_f)\right\vert^2G(\J_f),\,\,\,\,\,
\label{31}
\er
where the  $\b\b$-decay mode factors are
${\cal F}_{2\nu}=1$, and  ${\cal F}_{0\nu}=\langle m_{\nu}\rangle$,
with  $\langle m_{\nu}\rangle$ given   in natural units ($\hbar=m_e=c=1$).

\subsection {DCE Giant Resonances }\label{Sec2D}

In the present work, we will go a few steps further than what has been done in Refs.~\cite{Shi18,Men18,San18,San20},  comparing 
the NMEs in all final states that are reached by the $\b\b$-decay, and not only in the ground state,
and extending this study along the full strength distributions for the $0\nu\b\b$-decays and DCE reactions, up to the region of DCE giant resonances.
For this purpose, we introduce two new quantities:

1) The total   $0\nu\b\b^\mp$ strengths $S^{0\nu^\mp}$, 
and their energy distributions ${\textsf s}^{0\nu^\mp}(0^+_f)$, define as
\br
S^{0\nu^\mp}&=&\sum_{ f}|M^{0\nu^\mp}(0^+_f)|^{2}
\equiv\sum_{ f} {\textsf s}^{0\nu^\mp}(0^+_f),
\label{32}\er
as well as  the ratios
\br
R^{0\nu^\mp}(0^+_f) = {{\textsf s}^{0\nu^\mp}(0^+_f)}/{S^{0\nu^\mp}};
\label{33}\er %11 in pdf
and

2) The folded transition densities in $(A,Z\pm 2)$ nuclei
\br
{\mathscr S_{J\J}^{\{\mp 2\}}}(\E)=
\frac{\Delta}{\pi}\sum_{ f}\frac{ {\textsf s} ^{\{\mp 2\}}_{J\J}(\J_f)}{(\E-\E_f)^2+\Delta^2},
\label{34}\er
as a function of excitation energies
\be
\E_{f}=E^{\{\mp 2\}}_{0^+_f}-E^{\{\mp 2\}}_{0^+_1},
\label{35}\ee
of the final states $\ket{0^+_f}$, with the energy interval $\Delta=1$ MeV.  
% {\color{red} 

The transitions $S^{0\nu^\mp}_V$, $S^{0\nu^\mp}_{V_0}$, $S^{0\nu^\mp}_{A}$, and  $S^{0\nu^\mp}_{A_{1^+}}$, and the rates 
$R_V^{0\nu^\mp}(0^+_f)$,
$R_{V_0}^{0\nu^\mp}(0^+_f)$, $R_A^{0\nu^\mp}(0^+_f)$, and
$R_{A_1}^{0\nu^\mp}(0^+_f)$, as well as the corresponding    folded transition densities for  $M^{0\nu^{\mp}}_{V}$, $M^{0\nu^{\mp}}_{A}$, $M^{0\nu^{\mp}}_{V_0}$ and  $M^{0\nu^{\mp}}_{A_1}$, are defined in the same way.

The reaction folded strength densities meet the relationship 
\be
\int d\E{\mathscr S}^{\{\mp 2\}}_{J\J}(\E)=S^{\{\mp 2\}}_{J\J},
\label{36}\ee
 and the same holds for ${\mathscr S}_V(\E)$, and ${\mathscr S}_A(\E)$.

Note that, due to the neutron excess, in the medium and heavy nuclei, it occurs that 
 \br
 S^{\{{ -2}\}}_{J\J}\gg S^{\{{ +2}\}}_{J \J}, \; \mbox{and} \; S^{0\nu^-}_{V,A} \gg S^{0\nu^+}_{V,A}.
\label{37}\end{eqnarray}

\subsection {$Q$-Values }\label{Sec2E}

The $Q$-values for the $\beta\beta^-$-decay, and for the
$ee$-capture, are defined as
\br
Q_{\b\b^-}&=&\M(Z,A)-\M(Z+2,A),
\nn\\
Q_{ee}&=&\M(Z,A)-\M(Z-2,A),
\label{38}\er
where the $\M$'s are the atomic masses. 
When expressed by means  of  the energies $E_{0^+}^{\{0\}}$, and $E^{\{^\pm 2\}}_{0_1^+}$ that have been defined above, they read
\br
Q_{\b\b^-}&=&E_{0^+}^{\{0\}}-E^{\{-2\}}_{0_1^+}
\nn\\
Q_{ee}&=&E_{0^+}^{\{0\}}-E^{\{+2\}}_{0_1^+}.
\label{39}\er

%}

The above-mentioned observables are evaluated within different models of nuclear structure. 
Among them, the $pn$-QRPA is currently the most widely used nuclear model for estimating the NMEs
for the $\b\b $-decays to the ground state ${0_1^+}$
\footnote{
The SCE QRPA model was developed by Halbleib and Sorensen 
in  1967~\cite{Hal67} to describe the  $\b$-decay.
 Later, in the 1990s, averaging over two  $\b $-decays, this  QRPA was adapted for $\b\b $-decay calculations \cite{Vog86,Civ87,Tom87,Eng88,Hir90,Hir90a,Hir90b,Sta90,Krm92,Krm94}. 
The unfavorable aspects of the $pn$-QRPA model are that:
1) It cannot describe  $\b\b $- decays to excited states, and
 2) It does not allow the evaluation of the total  strengths \rf{2}, 
 and therefore neither the sum rules \rf{5} nor the 
 ratios $R^{\{\mp 2\}}_{J} (\J^+_f)$ given by \rf{33}, 
cannot be discussed either }.
 We have recently 
developed a new nuclear model for the DCE processes, based on the Quasiparticle Tamm-Dancoff 
Approximation (QTDA) for the $pn$ and $2p2n$ excitations~\cite{Fer20}. It is a natural extension of the 
original QRPA~\cite{Hal67}, and will be labeled DCEQTDA %called Double Charge-Exchange Model (DCEQTDA ). 
\footnote{This model has been proposed, and applied in its particle-hole limit 
for $\b\b $-decay in $^{48}$Ca, long years ago~ \cite{Krm05}.}.
 The graphical representation of the two models is given in~\cite[Fig. 1]{Fer20}.
The difference between them is cardinal, and has very important consequences, 
both analytical and numerical.
Below, we present the corresponding formulation within the DCEQTDA.

\section{\bf DCEQTDA Nuclear Model}
\label{Sec3}

% {\color{blue} 
The basic assumption in the DCEQTDA  is that the nuclei $(A,Z), (A,Z\pm1)$ 
and $(A,Z\pm2)$ can be represented, respectively, as the BCS vacuum 
in %{\color{blue} 
the initial nucleus %}
, and the excitations
of two $pn$ and four $2p2n$ quasiparticles  in this vacuum. The resulting 
eigenvalue problem is discussed in detail in our previous work \cite{Fer20}, where the model is labeled as $2p2n$-QTDA.

In the present work, we only deal with the resulting wave functions
\br
 \ket{J_i}&=&\sum_{pn}X_{pnJ_i}\ket{pnJ},
\nn\\
 \ket{\J_f}&=&\sum_{p_1p_2n_1n_2}\X_{{p_1p_2J_{p},n_1n_2J_{n}};\J_f} \ket{p_1p_2J_{p},n_1n_2J_{n};\J},
%\nn\\
\label{40}\er
and the corresponding energies $\w_{J_i}$ and $\Omega_{\J_f}$, 
for the intermediate and final nuclei, respectively.  
Here
\br
\ket{pnJ}&\equiv&A^\d  (pnJ) 
\ket{BCS},
\nn\\
\ket{p_1p_2J_{p},n_1n_2J_{n};\J}
&\equiv&\left[A^\d  (p_1p_2J_{p}) 
A^\d (n_1n_2J_{n})\right]_{\J}
\ket{BCS},
\label{41}\er
are, respectively, 
the intermediate proton-neutron two quasiparticle states, and the final two protons two neutron states, being
\br
A^\d(abJ)&=&N(ab)[a^\dag_aa^\dag_b]_{J}, \hspace{.4cm}
%\nn\\
N(ab)=\frac{1}{\sqrt{1+\delta_{ab}}},
\label{42}\er
 normalized two-quasiparticle states.
 and
$a^\d_a$ and the single-quasiparticle creation
 operators, defined by the Bogoljubov
transformation~\cite[Eqs.~(13.10)]{Suh07}.
 The BCS ground state is defined as $a_k\ket{BCS}=0$.

The amplitudes $X_{pnJ_i}$, and $\X_{{\rm p_1p_2J_{12},n_1n_2J'_{12}};\J^{+}_f}$, as well as the energies $\w_{J_i}$ and $\Omega_{\J_f}$, 
are obtained by diagonalizing the Hamiltonian
 \cite{Pal66,Raj69}
\br
&&H = H_0+H_{pn}+H_{nn}+H_{pp},
\label{43}\er
%{\color{red}  
%in the basis \rf{41}, \ie by solving the eigenvalue problems}
%{\color{blue}  
in both bases \rf{41}, \ie solving  two different eigenvalue problems
\br
&&H\ket{J_i}\equiv (H_0+H_{pn}) \ket{J_i}  =w_{J_i}\ket{J_i},
 \nn\\
&&H\ket{\J^{+}_f}\equiv (H_0+H_{pn}+H_{nn}+H_{pp}) \ket{\J^{+}_f}    =\Omega_{\J^{+}_f}\ket{\J^{+}_f},
\label{44}\er
the first for the intermediate nucleus, obtaining the amplitudes  $X_{pnJ_i}$, and the second for the final nucleus, obtaining the amplitudes
$\X_{{\rm p_1p_2J_{12},n_1n_2J'_{12}};\J^{+}_f}$.
%}{\color{red}  where} 
%{\color{blue}  
Here
%}
$H_0=\sum_a E_aa_a^\dag a_a$
is the independent-quasiparticle Hamiltonian, and
$H_{nn}$, $H_{pp}$, and   $H_{pn}$ are, respectively, neutron-neutron, proton-proton, 
and proton-neutron residual interactions among quasiparticles.
Obviously, $H_{nn}$, and $H_{pp}$ are not involved in the first of these two equations, since in \rf{44} $H_{nn}\ket{pnJ_i}=H_{pp}\ket{pnJ_i}=0$.
Details on the evaluation of matrix elements of $H$ can be found
in References~\cite{Pal66,Raj69}.
%}{\color{blue}  

As discussed in Ref. ~\cite{Fer17}, in the $pn$-QRPA calculation of the $\b\b$-decay is based on  the next approximations: 

i) In  \rf{14}, 
\brn
\Bra{\J^+_f}\O^{\pm}_J\Ket{J^{+}_i}\cong
\Bra{\bar J^{+}_i}\O^\mp_J\Ket{\bar 0^{+}}=\sum_{pn}\rho^\mp(pn\bar J_i^\pi) W_{J0J}(pn),
\ern
 where $\ket{\bar 0^{+}}$ and $\ket{\bar J^+_i}$ are, respectively, the ground states in the final nucleus, and the $pn$ excitations on it, and 

\ ii) In \rf{19}, 
\brn
E_{0^+}^{\{0\}}-E^{\{-2\}}_{0_1^+}\cong Q_{\b\b^-}, \hspace{.1cm} E_{0^+}^{\{0\}}-E^{\{+2\}}_{0_1^+}\cong Q_{ee},
\ern
being $Q_{\b\b^-}$, and $Q_{ee}$ taken from the experimental data for the
$Q$-values (see \rf{39}). 

That is,
one needs to compute the one-body F/GT transition matrix elements starting from both initial and final nuclei, and then match the intermediate states by computing the overlaps of their wave functions \cite{Aun96}. 

%{\color{blue} 
As a consequence, in the standard $pn$-QRPA \cite{Civ87} calculations, in addition to solving the BCS equation for the initial nucleus $(A,Z)$, and working with the initial occupation coefficients $(v_p,v_n)$
\footnote{The u’s and v’s in
the initial nucleus are determined 
under the constraints
$\sum_{j_p}(2j_p+1) v_p^2=Z$, and
$\sum_{j_n}(2j_n+1) v_n^2=N$, 
where $Z$ and $N$ are the number of protons and neutrons,
respectively, in the initial nucleus.
}, as we do in DCEQTDA, we must also solve the BCS equation for the final nucleus $(A,Z+2)$, or $(A,Z-2)$, and also deal with the final  occupation coefficients $(\bar v_p,\bar v_n)$. Moreover,
in the non-standard $pn$-QRPA only one BCS equation is solved, but for intermediate nucleus $(A,Z+1)$, or $(A,Z-1)$ (See Ref. \cite{Fer17}, and references therein). 
%}

In the DCEQTDA, %{\color{red}, conversely, } 
the wave functions of the final nucleus $ \ket{\J_f}$ are constructed directly on top of the wave function of the initial nucleus in terms of the $2p2n$ quasiparticle excitations. These wave functions, which do not exist in the $pn$-QRPA,  are then used to evaluate the two-body NMEs of $\b\b$-decay.

The $\b^-$-decay densities are evaluated 
from Eq.~\rf{3} and Eq.~\rf{4}, with the results 
\br
&&\varrho^-(p_1n_1;J_i)
=\sum_{pnJ_{p}} u_{p_1}v_{n_1} X_{p_1n_1J_i} ,
%\hat{J}^{-1}\Bra{{J_i}}(c^{{\dagger}}_{p_1}c_{\bar{n}_1})_{J}\Ket{0^+},
\label{45}\er
and
\br
&&\varrho^-(p_2n_2;J_i,\J^+_f)=\hat{J}\hat \J_f
  \sum_{pnJ_{p}J_{n}} (-)^{J_p+J_n}\hat{ J_p}\hat{J_n}
\nn\\
&\x&N(nn_2)N(pp_2)\X _{{\rm p p_2J_{p},nn_2J_{n}};\J_f}
\nn\\
&\x&{\bar P}(nn_2 J_n){\bar P}(pp_2J_p)
\ninj{p}{p_2}{J_{p}}{n}{n_2}{J_{n}}{J}{J}{\J} u_{p_2}v_{n_2} 
X_{pnJ_i},
%\nn\\
\label{46}\er
where  $\hat{J}=\sqrt{2 J+1}$, and
\br
N(ab)&=&(1+\delta_{ab})^{-1/2},
\label{47}\er
is the two particle normalization factor. %, and 
The operator
\br
{\bar P}(p_1p_2J)=1-(-)^{p_1+p_2-J}P(p_1\leftrightarrow p_2),
\label{48}\er
takes into account the Pauli Exclusion  Principle 
by exchanging among valence quasiparticles
 $p_1$ and $p_2$, and acts only on the right hand side \cite{Raj69}.
%the reason why the chemical potentials appear in Eq.(41). In order to make the given.
\footnote{
\brn
&&\varrho^-(p_2n_2;J_i,0^+_1)
\nn\\
&=&\sum_{pnJ_{p}} \hat{ J_p}
N(nn_2)N(pp_2)\X _{{\rm p p_2J_{p},nn_2J_{p}};0_1}
\nn\\
&\x&{\bar P}(nn_2 J_p){\bar P}(pp_2J_p)
\nn\\
&\x&(-)^{p_2+n_2+J_p+J}
\sixj{p}{p_2}{J_{p}}{n_2}{n}{J}
  u_{p_2}v_{n_2} X_{pnJ_i},
\ern}
%\newpage

The corresponding  $\b^+$-decay   densities 
density matrices $\varrho^+(p_1n_1;J_i)$, and 
$\varrho^+(p_2n_2;J_i,\J^+_f)$
are obtained from Eq.~\rf{49} 
and Eq.~\rf{50}, respectively,  after making 
$u_{p_1}v_{n_1}\go u_{n_1}v_{p_1}$,
and $u_{p_2}v_{n_2}\go u_{n_2}v_{p_2}$.

The two-body density matrix $\varrho^-(p_1n_1p_2n_2;J_i,\J_f)$, and $\bar\varrho^-(p_1n_1p_2n_2;J\J_f)$
 evaluated, respectively, from 
 Eqs. \rf{15} and  \rf{17} are
%\newpage
\br
&&\varrho^-(p_1n_1p_2n_2;J_i\J^+_f)=\hat{J}\hat \J_f
\nn\\
&\x&  \sum_{pnJ_{p}J_{n}} (-)^{J_p+J_n}\hat{ J_p}\hat{J_n}N(nn_2)N(pp_2)
\nn\\
&\x&\X _{{\rm p p_2J_{p},nn_2J_{n}};\J_f}
{\bar P}(nn_2 J_n){\bar P}(pp_2J_p)
\nn\\
&\x&
\ninj{p}{p_2}{J_{p}}{n}{n_2}{J_{n}}{J}{J}{\J} u_{p_2}v_{n_2} 
X_{pnJ_i}u_{p_1}v_{n_1} X_{p_1n_1J_i},
\label{49}\er
 and
\br
&&\bar\varrho^-(p_1n_1p_2n_2;J\J_f)=\hat{J}\hat \J_f
\nn\\
&\x&  \sum_{J_{p}J_{n}} (-)^{J_p+J_n}\hat{ J_p}\hat{J_n}N(n_1n_2)N(p_1p_2)
\nn\\
&\x&\X _{{\rm p_1p_2J_{p},n_1n_2J_{n}};\J_f}
{\bar P}(n_1n_2 J_n){\bar P}(p_1p_2J_p)
\nn\\
&\x&
\ninj{p_1}{p_2}{J_{p}}{n_1}{n_2}{J_{n}}{J}{J}{\J} u_{p_2}v_{n_2} u_{p_1}v_{n_1}.
\label{50}\er

\begin{figure}[t]
\centering
\includegraphics[width=8.5 cm,height=6.5 cm]{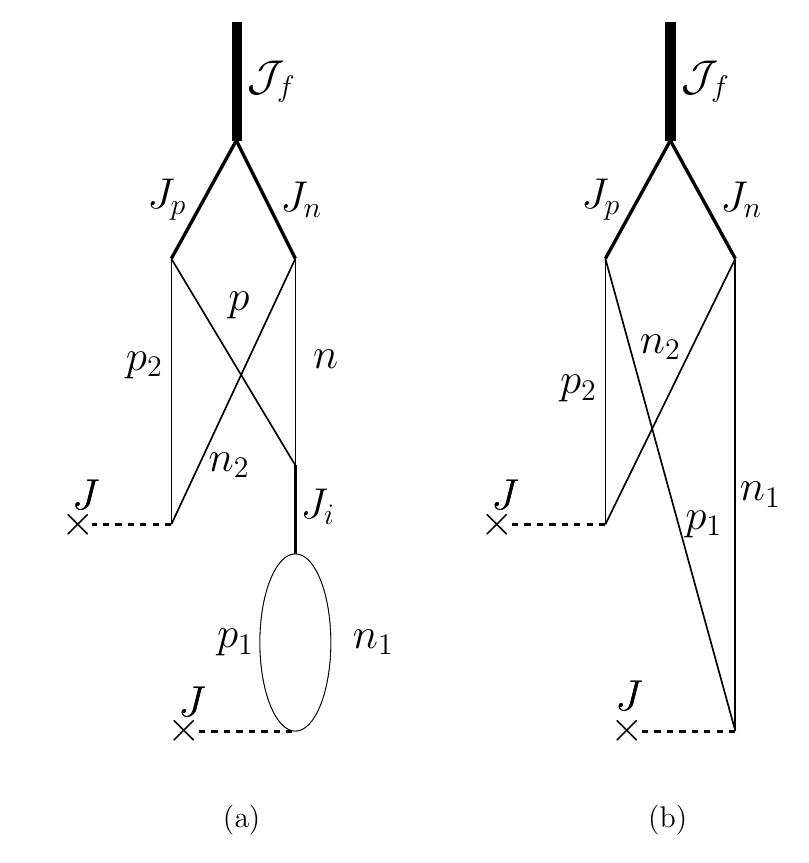}
\caption{\label{F1}
Two-body transition densities $\varrho(p_1n_1p_2n_2;J_i\J_f)$, 
given by ~\rf{49},
and  used in the evaluation of  $M^{2\nu}(\J_f)$,   
and  $\bar \varrho(p_1n_1p_2n_2;JJ_f)$, given by ~\rf{50}, 
and used to calculate $M^{0\nu}(0^+_f)$ within the CA  and to $M_J(\J_f)$,
are shown in panels (a) and (b), respectively. The dashed lines indicate the single $\b$-decays.
The first  $\b$-decay  is activated in the initial state, and the second in the intermediate state.  
The five vertices of these diagrams correspond to five of the six angular momentum couplings 
in the symbol 9j in Eqs. \rf{49} and \rf{50}. The sixth coupling $(JJ)\J$
corresponds to the three unconnected lines.
 }
\end{figure}

The graphical representations of densities 
$\varrho(p_1n_1p_2n_2;J_i^\pi\J^+_f)$, and 
$\bar\varrho(p_1n_1p_2n_2;J^\pi\J^+_f)$
are exhibited, respectively,  in panels (a),  and (b) of  Fig.~ \ref{F1}.

%%%%%%%%                                F1

%\newpage
The energies of the intermediate state $\ket{J_i }$ and the final state $\ket{\J_f }$, 
relative to the initial state $\ket{0^+}$, are  \cite{Fer20,Lan64,Rin80}
\begin{eqnarray}
E_{J_i }^{\{\mp 1\}}-E_{0^+}^{\{0\}}&=&
\w_{J_i}\pm\lambda_p\mp\lambda_n,
\nn\\
E^{\{\mp 2\}}_{\J_f}-E_{0^+}^{\{0\}}&=&
\Omega_{\J_f}\pm2\lambda_p\mp 2\lambda_n,
\label{51}\end{eqnarray}
where $\lambda_p$  and $\lambda_n$ are 
the proton and neutron chemical potentials, which are omitted in many $pn$-QRPA calculations, 
But, the energy denominator \rf{19}, which becomes
\br
{\cal D}_{J_i,\J_f}
&=&\w_{J_i}-\frac{\Omega_{\J_f}}{2},
\label{52}\er
being the same for $\b\b^\mp$-decays, does not depend of them.
%%%%%%%%%%%%%%%%%%%%%%%%%%%%%%%%

Contrarily, the $Q$-values, for the $\beta\beta^-$-decay and for the
$ee$-capture, in the present model, are given by 
\br
Q_{\b\b^-}&=&-\Omega_{0_1^+}-2(\lambda_p- \lambda_n),
\nn\\
Q_{ee}&=&-\Omega_{0_1^+}+2(\lambda_p- \lambda_n).
\label{53}\er
These are the windows of excitation energies where $\b\b$-decay 
and the ee-capture can be observed. Note that their difference 
\br
\Delta Q\equiv Q_{ee}-Q_{\b\b^-}=4(\lambda_p-\lambda_n),
\label{54}\er
depends solely on the mean field, i.e. on the difference between the chemical potentials.
\footnote{
It might be interesting to note that the chemical potentials $\lambda_p$, and $\lambda_p$ in Eqs. (51), are the same as in \cite[Eq.(2.38)]{Fer20}, \cite[page 58]{Lan64}, and  \cite[pages 235-237]{Rin80}. Moreover,   while they play no role in the $pn$-QRPA, in the DCEQTDA they allow us to evaluate the $Q$-values \rf{53}. which agree with the experimental data, as seen in  \cite[[Table Vi]{Fer20}, and here in Table I.}

%\footnote{where the reason why the chemical potentials appear in Eq.(41). In order to make the given.}
%%%%%%%%%%%%%%%%%%%%%%%
%\newpage

\section{\bf Numerical Results and Discussion}
\label{Sec4}
%{\color{blue} STOP}

As mentioned above, the main objective of this work is the evaluation of the $0\nu\b\b$ NMEs, within the DCEQTDA model, both analytically and numerically. The latter is done in Section \ref{Sec4C} by making use of Eqs.  \rf{22}, \rf{24}-\rf{27}, and \rf{53}.  Previously, in Section \ref {Sec4A} we explain how the numerical calculations are performed within this new model, and in Section \ref{Sec4B} we show the results for the DCE observables measured so far in $^{76}$Ge and $^{124}$Xe. This is for: 
the DCE $Q$-values  
$Q_{\b\b^-}$, and 
$Q_{ee}$, and for the excitation energies $\E_{\J^+_f}$ in the final nuclei, using, respectively, Eqs. \rf{39}, and \rf{51}, and for (ii) the NMEs $M^{2\nu^\mp}_{J}(\J_f)$, employing  \rf{16}, and \rf{18}. We proceed similarly with the remaining nuclei.

\subsection {\bf  Residual Interaction and Single Particle Energies}
\label{Sec4A}
We describe the residual interaction in 
\rf{43}, i.e in $H_{pn}$, $H_{nn}$, and $H_{pp}$ with the same $\delta$-force 
\be
V=-4\pi({\textit v}^sP_{s}+{\textit v}^tP_{t})\delta(r)\hspace{.5 cm}  
\mbox{MeV$\cdot$fm$^{3}$},
\label{55}\ee
because of its simplicity, and because in this way a comparison can be made between the present DCEQTDA, and our previous $pn$-QRPA calculations of the $0\nu$ NMEs ~[Fig. 3]\cite{Fer17}.

%for example PRC 107, 044316 (2023), go further e

To highlight the differences between the two models, 
we will use here the same  parameterization that 
was used in the $pn$-QRPA:

1) The pairing strengths for protons and
neutrons, ${\textit v}^s_{\rm pair}({p})$ and
${\textit v}^s_{\rm pair}({n})$ were obtained from the
fitting of the corresponding experimental pairing gaps.

2) The isovector  (${\textit v}^s$) and isoscalar  (${\textit v}^t$)
parameters within the particle-particle (pp) and particle-hole (ph) channels,
as well as the ratios $s={\textit v}_{\rm pp}^s/{\overline v}^s_{\rm pair}$,
and  $t={\textit v}_{\rm pp}^t/{\overline v}^s_{\rm pair}$,  with
${\overline v}^s_{\rm pair}=({{\textit v}^s_{\rm pair}({p})+{\textit v}^s_{\rm pair}({ n})})/2$,
were fixed in the same way as in our QRPA  calculations~\cite{Fer17,Fer20} (see ~\cite[Fig. 1]{Fer17}.
That is,  they are determined from the condition that the strengths $S^{{\{+ 1\}}}_{ J}$ become minimal, 
with the result $s=s_{sym}=1$, and $t=t_{sym}   \gtrapprox 1$, and which is named    Partial SU(4) Symmetry Restoration (PSU4SR).

In the present case we have $t_{sym}=1.34$ for  $^{76}$Ge, 
and $t_{sym}=1.42$ for  $^{124}$Xe.   This parametrization will be labeled as P1.
We have found it convenient to show in the case of $^{76}$Ge also the results 
for the value of the parameter $t$ that reproduces the measured value of 
the ground state $2\nu$ NME. This is frequently done in the literature~\cite{Sen13}, 
with the result here $t=1.86$, which we will label as P2.

All NMEs were evaluated with  the measured value
$g_{\sss A}=1.27$~\cite{Ber12}. 
The $\b\b$-decay  half-lives were calculated from \rf{31}, and 
all  kinematics
factors $G_{2\nu}(\J^+_f)$, and  $G_{0\nu}(\J^+_f)$ 
were taken  from~ \cite{Mirea15}, except for $G_{2\nu}(2^+_2)$ in $^{76}$Ge
that was  found in ~\cite{Suh98}.

%{\color{blue}
Although the most recent pn-QRPA calculations \cite{Jok23} include many single-particle levels ($18$ for $^{76}$Ge, and $26$ for $^{130}$Xe),
 we choose single-particle spaces such that they satisfy the sum rules \rf{10} and \rf{11}, for which they must have a sufficiently large number of states, these being the same for protons and neutrons, and always including the two spin-orbit partners. At the same time, they must be 
 numerically tractable in DCEQTDA.
 The  single-particle energies were obtained from the Wood-Saxon 
potential with the standard
parameterization ~\cite{Auo96}.
%}

%{\color{red}
%In order to satisfy the sum rules \rf{10} and %\rf{11} it is necessary to work 
%with a sufficiently large number of single-%particle states, which have to be the 
%same for protons and neutrons, and spin-orbit %partners must always be included. 
%{\color{red}The corresponding single-%particle}  Their   energies were obtained %from the Wood-Saxon energies were obtained %from the Wood-Saxon 
%potential with the standard
%parameterization ~\cite{Auo96}. }

Thus, we  use 9  single-particle levels
(2d$_{3/2}$,    1g$_{7/2}$     3s$_{1/2}$,     2d$_{5/2}$,   1g$_{9/2}$,       
2p$_{1/2}$,        1f$_{5/2}$,        2p$_{3/2}$,  and      1f$_{7/2}$)  in  $^{76}$Ge,  
and  7   levels ( 1h$_{7/2}$,  
1h$_{9/2}$,  2d$_{3/2}$,     1g$_{7/2}$,      
3s$_{1/2}$,     2d$_{5/2}$,   and 1g$_{9/2}$) in  $^{124}$Xe. The number  of resulting  
four quasiparticle states  $\ket{\J^+}$, defined 
in Eq.~\rf{40}, are $2045$ states  $\ket{0^+}$, and $8456$ states  $\ket{2^+}$ in $^{76}$Se, 
and $1146$ states  $\ket{0^+}$, and $4918$ states  $\ket{2^+}$ in $^{124}$Te. 
Some calculations have been performed on this nucleus  with 9 levels, 
including also the states 1f$_{7/2}$ and 1f$_{5/2}$.
The number  of resulting  
four quasiparticle states  $\ket{\J^+}$, define in Eq.~\rf{40}, are $2045$ states  $\ket{0^+}$, and $8456$ states  $\ket{2^+}$ in $^{76}$Se, 
and $1146$ states  $\ket{0^+}$, and $4918$ states  $\ket{2^+}$ in $^{124}$Te. 
%Some calculations have been performed on this %nucleus  with 9 levels, 
%including also the states 1f$_{7/2}$ and 1f$_{5/2}$.

\subsection {\bf  Measured DCE Observables}
\label{Sec4B}

%%%%%%%%T1
\begin{table}[h]
	\centering
	\caption{\label{T1}Calculated and experimental $Q$-values (in MeV)  
		for $^{76}$Ge. The parameterizations P1, and P2 are explained in the text.}
	\begin{small}
		\newcommand{\cc}[1]{\multicolumn{1}{c}{#1}}
		\renewcommand{\tabcolsep}{1.3pc} % enlarge column spacing
		\renewcommand{\arraystretch}{1.} % enlarge line spacing
		\begin{tabular}{cccc}
			\hline\hline
			Par/Exp &  $Q_{\b\b^-} $
			& $Q_{ee} $& $\Delta Q$\\
			\hline
			$^{76}$Ge &     &        & \\
			P1 & 1.239 &    -8.240  &-9.479 \\
			P2& $1.314$ &$ -8.164$ &    -9.478 \\
			Exp.~\cite{NNDC} &2.039  &    -10.910  &-12.95  \\
			\hline
			$^{124}$Xe& &&         \\
			P1 &-9.403 &    4.071 &13.474 \\ 
			Exp.~\cite{NNDC} &-8.572 &    2.864&11.436   \\
			\hline\hline
		\end{tabular}
	\end{small}
\end{table}
%%%%%%%%%%%%%%%

First, we will focus on the DCE observables  that have been experimentally measured, and will serve to test the nuclear model used.
This will be done for the  $\b\b$-decay
in $^{76}$Ge, and  for the $ee$-capture in $^{124}$Xe,  analyzing:

a) The $Q$-values, defined in \rf{39},
 that  were evaluated from \rf{53}, and
 are given in Table \ref{T1}. 
The agreement between the data and the calculations is quite reasonable.
It should be noted that we obtain that $Q_{\b\b^-}$ is positive and $Q_{ee}$ negative for $^{76}$Ge, and opposite for  $^{124}$Xe, which, although expected, is not a trivial result.

b) The  excitation energies $\E_{\J^+_f}$ in the final nucleus,  defined by  \rf{51}, and the   $2\nu$
NMEs   for   $\J^+_f=0^+_{1,2}$, and  $2^+_{1,2}$, both are listed in Table \ref{T2}. 
All energies are reasonably well reproduced, except $\E_{2^+_1}$ in $^{124}$Te,  
which turned out to be negative. 
%\footnote{
%{\color{red}
% The DCEQTDA prediction of a 2$^+$ state for ground state inf $^{124}$Te is obviously incorrect, since in nature the $0^+$ state is always the ground state for even-even nuclei. However, in theoretical calculations this only necessarily occurs in the case of having two identical particles, for which there is only one state for each total angular momentum, and when they interact with each other with the pure pairing force.
% Here, we work with the $\delta$-force  \rf{55} , where the pairing interaction is dominant, but other multipoles are also affected. Moreover, as mentioned above, in the $^{124}$Te case, we deal with 4 non-identical particles, where there are $1146$ $\ket{0^+}$ states, and $4918$ $\ket{2^+}$ states, which makes a big difference.
% The collective (vibrational) degrees of freedom play an important role in even-even Te isotopes, and should  be taken into account in any realistic calculation.  
%(See for example Ref. \cite{Krm77}.)
%}}

%{\color{blue}
The DCEQTDA prediction of $2^+$ state as ground state in $^{124}$Te is obviously incorrect, since in nature   the ground  state in even-even nuclei is always the $0^+$ state.  In theoretical calculations this only necessarily occurs in the case of having two identical particles, for which there is only one state for each total angular momentum, and when they interact with each other with the pure pairing force.
 Here we work with the $\delta$-force  \rf{55}, where the pairing interaction is dominant, but other multipoles  are also involved. Moreover, as already mentioned, in the case of $^{124}$Te, we deal with 4 non-identical particles, where there are $1146$ $\ket{0^+}$ states, and $4918$ $\ket{2^+}$ states, which makes a big difference. The collective (vibrational) degrees of freedom play an important role in the even isotopes of Te, and must be taken into account in any realistic calculation.  (See, for example, Ref. \cite{Krm77}).
 %}

c) The corresponding half-lives $\tau_{2\nu}$ are also shown, 
and compared with the experimental data \cite{Agos20,Bar20,Apr19}. 
All our computed half-lives ${\tau_{2\nu}}(\J^+_f)$ are consistent with the measurements.

For comparison, Table \ref{T2} further shows the calculations in $^{76}$Ge performed within: i) the Multiple-Commutator Model 
(MCM) \cite{Pir15}
\footnote{The MCM is in fact a superposition of two different nuclear models, each with its own parameterisation. One is the charge-exchange pn-QRPA, used to describe the intermediate states $\ket{J_i}$,
and the other is a charge-conserving QRPA, used to describe the excited final states $\ket{J^+_f}$.}, the Shell Model (SM) \cite{Kos22}, as well as the calculations 
in $^{124}$Xe performed within: i) 
MCM ~\cite{Pir15}, and ii) the Effective Theory (ET), and 
the SM \cite{perez2019two}.

%\ne

%%%%%%T2
\begin{sidewaystable}
%\begin{table*}[h]
	\caption{
		Calculated and experimental excitation
		energies $\E_{\J_f}$ in $^{76}$Se,  the  NMEs  $M^{2\nu}(\J^+_f)$, 
		and  half-lives ${\tau_{2\nu}}(\J^+_f)$
		for the decays of $^{76}$Ge, and $^{124}$Xe  to $\J^+_f=0^+_{1,2}$, and $2^+_{1,2}$
		states in $^{76}$Se, and $^{124}$Te, respectively,  with parameterizations P1 and P2,
		and with $g_{\sss A}=1.27$ ~\cite{Ber12}. This table also shows the calculations 
		in $^{76}$Ge performed within: i) the Multiple-Commutator Model 
		(MCM),
		%{\color{red}, }
		~\cite{Aun96}, and ii)  
		the Shell Model (SM) \cite{Kos22}, as well as the calculations 
		in $^{124}$Xe performed within: i) 
		MCM ~\cite{Pir15}, and ii) the Effective Theory (ET), and 
		the SM \cite{perez2019two}, 
		where all NMEs are multiplied by $(g_{\sss A}=1.27)^2$.
		The experimental % {\color{blue} 
			ground state
			%} 
		${\tau_{2\nu}^{ee}}(\J^+_f)$ is from Ref. ~\cite{Apr19}. 
		The measured half-lives  in $^{76}$Se are  from~ \cite{Bar20} for the ground state, 
		and from~\cite{Arn21} for the excited states,
		and  the experimental  ${\tau_{2\nu}^{ee}} $ in $^{124}$Te is from Ref.~\cite{Apr19}.
		All $G$ factors are from Ref.~\cite{Mirea15}, except  that of the $2^+_2$ state in $^{76}$Se, 
		which is from Ref.~\cite{Suh98}.}
	\label{T2}
	\centering
	\begin{small}
		\newcommand{\cc}[1]{\multicolumn{1}{c}{#1}}
		\renewcommand{\tabcolsep}{2.0pc} % enlarge column spacing
		\renewcommand{\arraystretch}{1.} % enlarge line spacing
		\begin{tabular}{cccccc}
			\hline\hline
			$^{76}$Ge $\go$ $^{76}$Se & $\J^+_f$ & $0^+_1$ & $0^+_2$ & $2^+_1$  & $2^+_2$  \\
			\hline
			$\E$  (MeV)                   &     &          &    &    &      \\
			&   P1   &      $0.0$     & $2.13$    & $ 0.36 $     &  $1.98$      \\
			&   P2   &      $0.0$     & $2.13$    & $0.26$     &  $1.90$      \\
			&{ Exp.}  &      $0.0$     & $1.12$    & $0.56$     &  $1.22$      \\
			\hline
			$\vert M^{2\nu} \vert$ (n.u.)   &   & $\x 10^{3}$ & $\x 10^{3}$ &  $\x 10^{3}$ & $\x 10^{3}$   \\
			& P1   &      $56.4$      & $80.1$   &  $2.8$  & $5.2$      \\
			& P2   &      $103$      & $158$   &  $6.6$  & $12.3$      \\
			&SM     &  $97$   & $70$  &  $0.7$ & $1.8$         \\
			&MCM & $74$          & $363$          &$1$ &$3$  \\
			&{ Exp.}&      $107$      & $ < 147$       &$  < 58$       &$ < 873$          \\
			\hline
			%\multirow{5}{0.4 in}{${\tau_{2\nu}}$ (yr)}
			${\tau_{2\nu}}$ (yr)&&$\x 10^{21}$      & $\x 10^{23}$      & $\x 10^{23}$ & $\x 10^{24}$    \\
			& P1   &    $6.8$  &   $25$   & $ 3.3\cdot 10^{3}$&  $ 3.6\cdot 10^{4}$\\
			& P2   &    $2.0$  &   $6.5$  & $ 6.0\cdot 10^{2}$&  $ 6.5\cdot 10^{3}$ \\
			& Exp. &   $1.88$  &  $> 7.5$         &$> 7.7$  &$> 1.3$       \\
			\hline
			$^{124}$Xe $\go$ $^{124}$Te & $\J^+_f$ & $0^+_1$ & $0^+_2$ & $2^+_1$  & $2^+_2$  \\
			\hline
			$\E$  (MeV)&&        &   &    &     \\
			&   P1   &      $0.0$     & $1.74$    & $-0.16$     &  $1.62$      \\
			$$  &{ Exp.}  &      $0.0$     & $1.66$    & $0.60$     &  $1.32$      \\
			\hline
			%$G^{2 \nu ECEC}$         &          &    $1.72(-20)$     & $1.67(-22)$         & $1.38(-23)$ &       \\\hline
			%\multirow{4}{0.61 in}{$|M|^{2\nu}$  (n.u.) $\x 10^{3}$ }
			$\vert M^{2\nu} \vert$ (n.u.)   &   & $\x 10^{3}$ & $\x 10^{3}$ &  $\x 10^{3}$ & $\x 10^{3}$   \\
			& P1     &    $62$     & $46$         &  $7.1$               & $8.3$      \\
			&ET      & $18-66$   &  $3.2-80$  & $ 0.13-1.45$     &          \\
			&SM     &  $45-116$   & $8-16$  &  $0.17-0.37$ &          \\
			&MCM &     476       &   15        &   1.1      &  0.11        \\
			&{ Exp.}&  $57 \pm 10$      & { }       & {}        &          \\
			\hline
			%\multirow{4}{0.4 in}{${\tau_{2\nu}}$ (yr)}
			${\tau_{2\nu}}$ (yr)&&$\x 10^{22}$      & $\x 10^{25}$      & $\x 10^{28}$ &     \\
			&P1    &    $ 1.51 $           &   $ 0.28$               &    $0.14$   &       \\
			&ET     & $18.6-1.33$  &$57.8-0.92$ &$435-3.45$& \\
			&SM    & $2.87-0.43$  &$9.21-2.34$ &$231-52.6$&  \\
			&MCM & $0.026$          & $2.66$          &$5.99$ &  \\                            
			& Exp.&$ 1.8\pm0.5$     &                     &    &       \\
			\hline\hline
		\end{tabular}
	\end{small}
%\end{table*}
\end{sidewaystable}
%%%%%%%%%%%

%{\color{black} 
%
%The fact that the agreement with experiments is satisfactory for the $^{76}$Ge and $^{124}$Xe nuclei suggests the reliability of the nuclear model used.  The inclusion of the remaining nuclei in this discussion would not change that conclusion, but would make the article very tedious to read, due to the large amount of material included.

The fact that the agreement with experiments is satisfactory for the nuclei $^{76}$Ge and $^{124}$Xe, as well as for $^{48}$Ca, and $^{86}$Ru, which we have previously performed in Ref.  \cite{Fer20}, suggests the reliability of the nuclear structure model DCEQTDA.
Despite this, the question remains pertinent as to whether the inclusion of other nuclei would change the conclusions about its validity. The answer is the opposite, and that the more analysed cases, the more the validity of the DCEQTDA model is confirmed, as shown in Table \ref{T7}, which is delegated to the Appendix so that reading the article does not become too tedious, due to the large amount of material it includes.
%
% The inclusion of more nuclei in this discussion would %not change the conclusions, but would make the article %very tedious to read, due to the large amount of %material that was included.
%}
%\newpage
%%%%%%%%%%%%%%T3
\begin{sidewaystable}
%\begin{center}
%\begin{table*}[t]
\centering
\caption {Fine structure of $M^{0\nu}(0^+_1)$  for $^{76}$Ge, evaluated in the $pn$-QRPA and  DCEQTDA with the same parametrization P1 in both cases. The contributions of different intermediate-state angular momenta $J^\pi$
are listed for both parities $\pi=\pm$. The most relevant numbers are shown in bold type.}
\label{T3}
\bigskip
		\newcommand{\cc}[1]{\multicolumn{1}{c}{#1}}
       \renewcommand{\tabcolsep}{.7pc} % enlarge column spacing
       \renewcommand{\arraystretch}{1.} % enlarge line spacing
\begin{tabular}{c|ccccc|ccccccc|}
\hline
%\\
&&&$pn$-QRPA&&&&& DCEQTDA&&&\\
\hline
%\\
$J^\pi$ & $M^{0\nu}_V$& $M^{0\nu}_A$  &  $M^{0\nu}_M$ &$M^{0\nu}_{P}$ &$M^{0\nu} $
            & $M^{0\nu}_V$& $M^{0\nu}_A$  &  $M^{0\nu}_{M}$ &$M^{0\nu}_{P}$ &$M^{0\nu}$         \\
\hline
   0$^+ $& {\bf 0.020}&      0.000&   0.000&   0.000&    0.020     &{\bf0.145}&     0.000&   0.000&   0.000&    0.145\\
   1$^+ $&      0.000&{\bf-0.682}&   0.007&  -0.022&   -0.697     &     0.000&{\bf0.466}&   0.006&  -0.027&    0.445\\
   2$^+ $&      0.173&      0.255&   0.026&   0.000&    0.453     &     0.035&     0.047&   0.005&   0.000&    0.087\\
   3$^+ $&      0.000&      0.475&   0.028&  -0.172&    0.330     &     0.000&     0.077&   0.004&  -0.018&    0.064\\
   4$^+ $&      0.080&      0.121&   0.022&   0.000&    0.223     &     0.011&     0.014&   0.003&   0.000&    0.027\\
   5$^+ $&      0.000&      0.230&   0.024&  -0.104&    0.150     &     0.000&     0.026&   0.003&  -0.008&    0.021\\
   6$^+ $&      0.030&      0.049&   0.015&   0.000&    0.093     &     0.003&     0.004&   0.001&  -0.000&    0.008\\
   7$^+ $&      0.000&      0.097&   0.014&  -0.045&    0.066     &     0.000&     0.009&   0.001&  -0.003&    0.007\\
   8$^+ $&      0.009&      0.014&   0.005&   0.000&    0.028     &     0.001&     0.001&   0.000&   0.000&    0.002\\
   9$^+ $&      0.000&      0.056&   0.012&  -0.026&    0.042     &     0.000&     0.006&   0.001&  -0.002&    0.005\\
  \hline
 $\pi=+$ &      0.311&      0.613&   0.153&  -0.368&    0.708     &     0.194&     0.650&   0.024&  -0.057&    0.811\\
\hline
   0$^- $ &     0.000&      0.054&   0.000&  -0.029&    0.025     &     0.000&     0.009&   0.000&  -0.003&    0.006\\
   1$^- $ &     0.149&      0.227&   0.013&   0.000&    0.389     &     0.019&     0.023&   0.001&   0.000&    0.043\\
   2$^- $ &     0.000&      0.741&   0.025&  -0.170&    0.596     &     0.000&     0.148&   0.004&  -0.017&    0.135\\
   3$^- $ &     0.107&      0.230&   0.031&   0.000&    0.368     &     0.011&     0.028&   0.004&   0.000&    0.043\\
   4$^- $ &     0.000&      0.398&   0.033&  -0.144&    0.286     &     0.000&     0.048&   0.004&  -0.009&    0.043\\
   5$^- $ &     0.060&      0.132&   0.031&   0.000&    0.223     &     0.006&     0.015&   0.003&  -0.000&    0.024\\
   6$^- $ &     0.000&      0.201&   0.027&  -0.084&    0.144     &     0.000&     0.018&   0.002&  -0.004&    0.016\\
   7$^- $ &     0.028&      0.064&   0.023&   0.000&    0.115     &     0.003&     0.006&   0.002&   0.000&    0.012\\
   8$^- $ &     0.000&      0.050&   0.010&  -0.023&    0.037     &     0.000&     0.002&   0.000&  -0.000&    0.002\\
\hline
$\pi=-$   &     0.343&      2.098&   0.193&  -0.451&    2.183     &     0.039&     0.296&   0.021&  -0.034&    0.323\\
\hline
Total   &{\bf  0.654}&{\bf 2.711}&   0.346&  -0.819&{\bf 2.891 }  &{\bf0.234}&{\bf0.946}&   0.045&  -0.091&{\bf 1.133}\\
\hline\hline
\end{tabular}
%\end{table*}
%\end{center}
 \end{sidewaystable} 
%%%%%%%%%%%%%%%%%%%%%%%%%%%%%%%%%%%%%%%%%%%%%%%%%%%%%%%%%%%%%

\subsection {\bf $0\nu\b\b$  NMEs}
\label{Sec4C}
\subsubsection {\bf Anatomy of $0\nu\b\b$  NMEs}
\label{Sec4C1}
In Table \ref{T3} we show the contributions of different intermediate states $J^\pi$ to $M^{0\nu^-}(0^+_1)$ in $^{76}$Se (positive parities in the top panel and negative parities in the bottom panel, evaluated in the $pn$-QRPA and  DCEQTDA, with the same residual interaction \rf{55} and the parametrization P1, and using the same set of single-particle energies in both cases. 
\footnote {With the P2 parameterization, similar results are obtained in the DCEQTDA, but the $pn$-QRPA in this case collapses.}

The most relevant 
%{\color{blue} 
partial NMEs 
%}
in both nuclear models are shown in bold type,
 and are labelled as: 
 
  i) $M^{0\nu}_{V_{0^+}}$,  which  is the contribution of  intermediate states $J^\pi=0^+$  to $M^{0\nu}_{V}$,  and 
 
  ii) $M^{0\nu}_{A_{1^+}}$,  which  is the contribution of  intermediate states $J^\pi=1^+$  to $M^{0\nu}_{A}$.
It is important to note that the  $0\nu\b\b$ NMEs differ from the $2\nu\b\b$ and reaction NMEs, not only in radial dependence but also in genesis. Indeed, while %{\color{red} they}  {\color{blue}  
the first %} 
receive contributions from all intermediate states  $J^\pi=0^\pm,1^\pm, \cdots 9^\pm$, the latter are constructed only from the intermediate states $J^\pi=0^+ (M_V^{2\nu}$, and $M_{00}$), and $J^\pi=1^+ (M_A^{2\nu}$, and $M_{10}$).  
%\newpage

%{\color{red} 
%he most remarkable difference between the two calculations is that:
%i) while  in the  $pn$-QRPA  the contributions of the intermediate states with $J^\pi\ne 1^+$ are much  bigger than those of the $J^\pi= 1^+$ states, 
%and that they are of opposite sign,
%\footnote {Negative value of the $1^+$ contribution  ave been seen  in $^{96}$Zr, $^{100}$Mo, and $^{124}$Sn nuclei\cite{Kor07,Hyv15}. The %interference with the rest of the contributions reduces the magnitude of the final NME for these decays.  }
%ii)  in the DCEQTDA  the contributions of the intermediate states 
%with $J^\pi\ne 1^+$ are smaller than those of the $J^\pi= 1^+$ states, and that all spins contribute coherently. 
%The final result is that the total $0\nu\b\b$ NME $M^{0\nu}$ is factor of 3 lower 
%that the one obtained in the $pn$-QRPA calculation.

%To facilitate the comparison of the results of Refs. \cite{Sim08,Sen16} with ours, in Fig. \ref{F2} we show the total contributions of each intermediate state $J^\pi$ to $M^{0\nu^-}(0^+_1)$ in $^{76}$Se (positive parities in the upper panel and negative in the lower panel, evaluated in the same way as in Table \ref{T3}. From this comparison, we see that:
% a) Our  $pn$-QRPA agree both qualitatively and quantitatively with calculations performed by  \v Simkovic \etal ~\cite[Fig. 3]{Sim08},  based on the Bonn-CD nucleon-nucleon interaction, but employing the same nuclear model, while 
 
%b) The calculations performed in the DCEQTDA are consistent with the SM results obtained by  Sen’kov and Horoi \cite[Fig.3]{Sen16}.
%}

%{\color{blue}
Table \ref{T3} also shows that:

 i) In both nuclear models the Fermi (F), weak magnetism (M) and pseudoscalar (P) NMEs are small compared to the axial vector (A) ones.

 ii) As a consequence, in both nuclear models the contributions of the intermediate states with $J^\pi= 1^+$ dominate those with $J^\pi\ne 1^+$, and, in particular, in the $pn$-QRPA are negative, and therefore interfere destructively
  \footnote {Negative value of the $1^+$ contribution  ave been seen  in $^{96}$Zr, $^{100}$Mo, and $^{124}$Sn nuclei\cite{Kor07,Hyv15}. The interference with the rest of the contributions reduces the magnitude of the final NME for these decays.}.
Despite this, the final result is that the total NME $M^{0\nu}$
 in  the DCEQTDA is significantly lower than that obtained in the $pn$-QRPA calculation.
%}

 %{\color{blue} 
There are several previous QRPA calculations on the anatomy of NMEs as a function of intermediate states \cite{Civ05,Kor07,Sim08,Sim13,Hyv15,Jok18}, which were obtained using different single-particle spaces,
and various residual interactions.
They all lead to results that differ from each other, making it difficult to choose one to compare with ours. In this scenario, we will compare with those reported in the Ref. ~\cite{Hyv15}, done with two- nucleon interactions based on the Bonn one-boson-exchange G matrix. 
A similar study based on the shell model also exists ~\cite{Sen16}, which will also be used in our discussion.
  
\begin{figure}[th]
\centering
%\vspace{-6cm}
\hspace{-4.5cm}
\includegraphics[width=16.5cm,height=6.5cm]{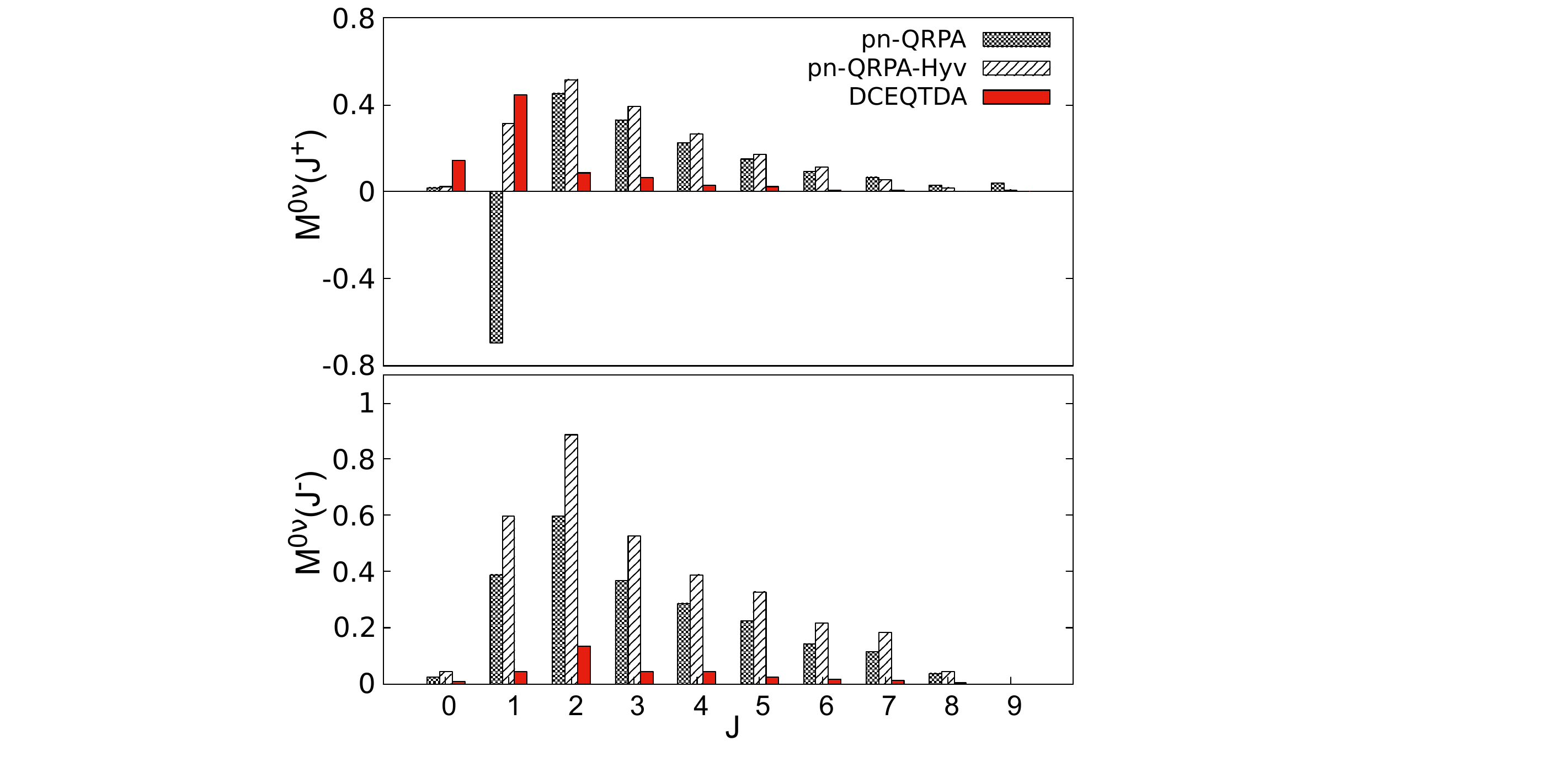}
\caption{\label{F2}
Contributions of different intermediate-state angular momenta $J^\pi$ to $M^{0\nu}(0^+_1)$ in$^{76}$Se within present  $pn$-QRPA, and  DCEQTDA calculations,  employing the same parametrization P2, and the $pn$-QRPA calculation done by  Hyv\"arinen, and Suhonen \cite{Hyv15}. Positive parities are shown in the upper panel and negative parities in the lower one. }
\end{figure}

To facilitate the comparison of the last two mentioned works  with ours, in Fig. \ref{F2} we show the
 total contributions of each intermediate state $J^\pi$ to $M^{0\nu^-}(0^+_1)$ in $^{76}$Se (positive parities in 
 the upper panel and negative parity in the lower panel), which are listed in column 6 (for pn-QRPA) and in column 11 (for DCEQTDA) of Table \ref{T3}. The same figure also shows the results obtained by Hyv\"arinen and  Suhonen
~\cite[Fig. 1]{Hyv15}.  From this comparison, we see that:
   
a) Except for the aforementioned sign of the intermediate $J^\pi= 1^+$ states, our $pn$-QRPA results agree qualitatively with the calculations performed in Ref.  ~\cite{Hyv15}, using the same nuclear model. However, their value $M^{0\nu}=5.26$ for the total NME is significantly larger than our result $M^{0\nu}=2.89$, which is probably due to the use of a more realistic nucleon-nucleon interaction than the one we use.

 b) On the other hand,  our DCEQTDA calculations are consistent with the results obtained by Sen'kov and Horoi \cite{Sen16} in the SM. 
In fact, in both cases, the most important contributions to $M^{0\nu}$ come from intermediate states $J^\pi= 1^+$, and  $J^\pi= 2^-$. These are, respectively $30\%$, and $15\%$  in Ref. \cite{Sen16}, and $39\%$, and $12\%$  in our case (see Table 3).
The difference in the total value of $M^{0\nu}$ (theirs $=3.374$, and ours $=1.133$) may come from the difference in both the residual interaction and the nuclear structure model used in the two calculations. 
%}

%{\color{red}
%From the results shown above it can be deduced that in a nuclear structure calculation the central role is played by the degrees of freedom included in the model, while the details of the employed nuclear force are of minor importance. In this case, it is the inclusion of four quasiparticles in the final nucleus that produces the big difference between the DCEQTDA and $pn$-QRPA results.
%}

%%%%%%%%%%%%%%%%%%
%%%%%%%%%%%%%

%%%%%%%%%%%%%%%%%%
\begin{figure}[ht]
	\centering
	\includegraphics[trim=2cm 2cm 0cm 0cm,width=11 cm,height=7cm]{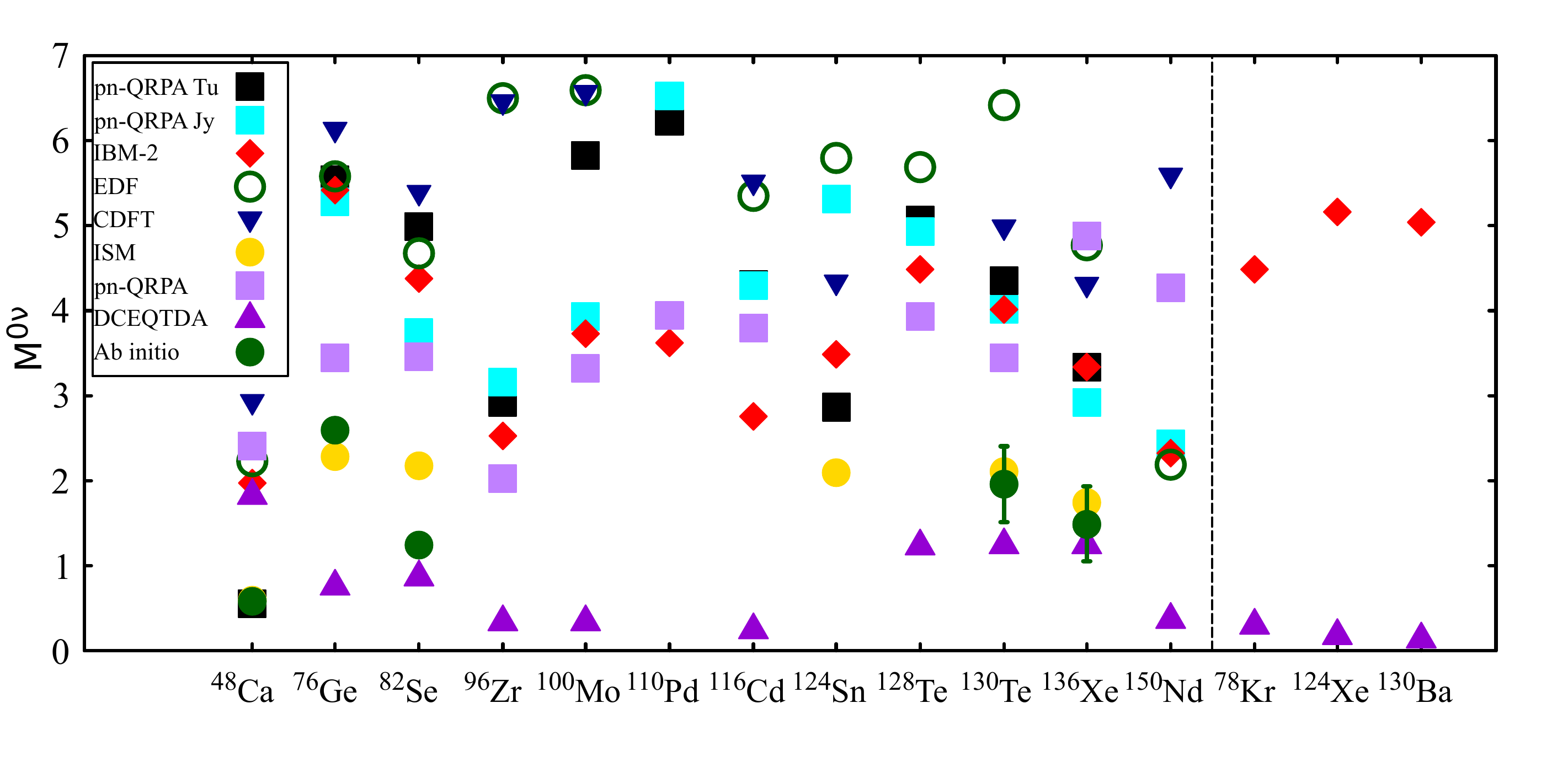}
	\caption{\label{F3}(Color online)   Comparison of calculated NMEs $M^{0\nu^-}(0^+_1)\equiv M^{0\nu}$, 
	within several nuclear structure models:
	i)   $pn$-QRPA by  T\"{u}bingen group (QRPA Tu)~\cite{Sim13}  ($g_{\sss A}=1.27$),
	ii)  $pn$-QRPA by Jyv\"{a}skyl\"{a} group(QRPA Jy) ~\cite{Hyv15} ($g_{\sss A}=1.26$), 
	iii) Interacting Boson Model (IBM-2)~\cite{IBM-2} ($g_{\sss A}=1.269$),
	iv)  Energy Density Functional Method (EDF)~\cite{EDF}  ($g_{\sss A}=1.25$),
	v)   Covariant Density Functional Theory (CDFT) \cite{Son14,Yao15}  ($g_{\sss A}=1.254$), 
	vi)  Interacting Shell Model (ISM)~\cite{Men09} ($g_{\sss A}=1.25$),   
	vii) The Ab  initio Calculations in  $^{48}$Ca, $^{76}$Ge, and  $^{82}$Se~\cite{Bel21}, in $^{130}$Te~\cite{Bel23}, and in $^{136}$Xe~\cite{Bel24}
	viii) The present  $pn$-QRPA and  DCEQTDA  results ($g_{\sss A}=1.27$).}
\end{figure}

\subsubsection {\bf Comparison of our results for $0\nu$ NMEs with other calculations}
%Comparison with  other calculations of $0\nu\b\b$  NMEs}
\label{Sec4C2}

In Fig. \ref{F3} we compare our $pn$-QRPA and DCEQTDA calculations for the ground-state $0\nu\b\b$, and  $0\nu ee$ NMEs, performed with the same residual interaction and identical nuclear model parameters. We also show the results of $pn$-QRPA calculations done with more realistic residual interactions, such as the Argonne V18 or CD-Bonn potentials~\cite {Sim13,Hyv15}, as well as the results of calculations with other nuclear models.

It can be deduced that, except for $^{48}$Ca, the new DCEQTDA model produces results that are definitely smaller than the other nuclear structure models; going from $0.12$ in $^{116}$Cd to $1.7$ in $^{48}$Ca for  $0\nu\b\b$  NMEs, and from $0.14$ in  $^{130}$Ba to $0.33$  in $^{78}$K for  $0\nu ee$ NMEs. {\it Ab-initio} calculations of $0\nu\b\b$  NMEs, ~\cite{Bel21,Bel23,Bel24} are going in the same direction, but we have no idea what the reason for this is! 

The fact that in the DCEQTDA  model says the NMEs are much more smaller in the $ 0\nu ee$-capture than in the  $ 0\nu\b\b^-$-decay,  is fully consistent with Eq. \rf{37}, but strongly disagrees with the $pn$-QRPA and IBM calculations, shown in  \cite[Table III]{Bar13}. 
  The relations  \rf{37} are completely independent of the nuclear model used and depend only on the neutron excess. 
However, in making the above comparison it should be noted that, as Fig. \ref{F4} indicates, most of the transition strengths $S^{0\nu^-}_{V,A}$ are concentrated in the region of giant resonances, far from the fundamental state of the final nucleus, and that such collective states do not appear in the strengths $S^{0\nu^+}_{V,A}$.
   
A more complete analysis of the   $2\nu\b\b$ NMEs is given in Table \ref{T7}, where the results for several nuclei are shown, 
in which double beta decay, and double electron capture processes take place.

\vspace{-0.1cm}
\subsubsection {\bf Relation between $0\nu\b\b$,  and Double Fermi and Double Gamow-Teller NMEs, within the DCEQTDA}
\label{Sec4C3}

The upper part of Table \ref{T4} presents the comparison between the calculations for:
i) The  $0\nu\b\b$ NMEs $M^{0\nu}_{V}$, $M^{0\nu}_{V_{0^+}}$, $M^{0\nu}_{A}$, $M^{0\nu}_{A_{1^+}}$, $M^{0\nu}_P$,  $M^{0\nu}_{M}$, and  $M^{0\nu} $, and
 ii) The absolute values of $M _{00}$, and  $M_{10}$,
 for $\J^+_f=0^+_{1,2}$ states  in $^{76}$Se,  with the P1 and P2 parameterizations. 
 
 From the discussion presented above in Section \ref{Sec4C1}, it is easy to be convinced that physically   $M _{00}$ corresponds to $M^{0\nu}_{V_{0^+}}$, since both are generated only from intermediate states $J=0^+$. The same is true for $M_{10}$ and $M^{0\nu}_{A_{1^+}}$, which are generated only from intermediate states  $J=1^+$. Therefore it is not correct to associate $M _{00}$ with  $M^{0\nu}_{V}$, nor $M_{10}$ with $M^{0\nu}_{A}$, as is often done.  In previous comparisons between $M _{10}$ and $M^{0\nu} $ ~\cite{Shi18,Men18,San18,San20}, the results obtained for them in different calculations, and with different nuclear models were used. 
 %In addition, the contributions of the NMEs: V, P and M to $M^{0}$ , which, although small, are significant, were omitted. REVISAR

 The fact that $0\nu\b\b$  and DCER (Double Charge Exchange Reaction) NMEs turn out to be of the same order of magnitude is a surprise in view of the large differences between them (radial and energy dependencies, finite nucleon size effects, and short-range correlations taken into account in the former, etc).
It should also be remembered that the nuclear radius $r_N$ was introduced into $M^{0\nu}$ by Doi~\cite{Doi93} some time ago in a rather arbitrary way, just to make it dimensionless (see \rf{28}). 

 Using the calculated NMEs in $^{76}$Ge, together with the measured half-lives:   
$\tau_{0\nu}(0^+_1)> 1.8 \cdot 10^{26}$ yr \cite{Agos20}, 
and $\tau_{0\nu}(0^+_2)> 7.5\cdot 10^{23}$ yr \cite{Arn21}, 
and the $G(0^+_{1,2})$ factors from \cite{Mirea15},
we get from \rf{31} %\rf{35} 
the lower limits for  $\langle m_{\nu} \rangle$, 
that are shown in the last column of Table \ref{T4}.

When compared with the calculation of the $2\nu\b\b $-decay from Table \ref{T2}, it is observed that the differences between the two results
for  $0\nu\b\b$ NMEs are relatively smaller. 
 On the other hand, as expected both current values of $M^{0\nu}(0^+_1)$ in $^{76}$Ge are
notably smaller than those obtained in our $pn$-QRPA calculation~\cite{Fer17}, which
 was $M^{0\nu^-}(0^+_1)= 3.19^{-0.24}_{+0.46}$.

 In the lower part of Table \ref{T4} are presented the $ 0\nu ee$-capture   NMEs for  the $\J^+_f=0^+_{1,2}$ final states  in $^{124}$Te,   evaluated with 7  and 9 single-particle 
levels, obtaining quite similar results.
In both cases, they are noticeably smaller than our $ 0\nu\b\b^-$-decay   
NMEs   for $^{76}$Se.

%%%%%%%TABLE 4
\begin{table}[t]
\begin{center}
\hspace{-.5cm}
\caption{
At the top is the comparison between 
 the NMEs $M^{0\nu^-}(0^+_f)\equiv M^{0\nu}$, and  $M^-_{J\J}(0^+_f)\equiv M_{J\J}$, 
 for the $0^+_f=0^+_{1,2}$ states in $^{76}$Se,  within the DCEQTDA, with parameterizations P1 and P2.
 The upper bounds of  $\langle m_{\nu} \rangle$ are evaluated from Eq. \rf{31}, and are given in units of eV. In doing so, we have used the measured half-lives: $\tau_{0\nu}(0^+_1)> 1.8\x 10^{26}$ yr \cite{Agos20}, 
 and  $\tau_{0\nu}(0^+_2)> 4\x 10^{23}$  yr  \cite{Arn21}.
The lower part of the table shows the same but only the  NMEs $M^{0\nu^+}(0^+_f)\equiv M^{0\nu}$ 
for electron capture in the $0^+_f=0^+_{1,2}$ states of $^{124}$Te, obtained with 7, 
and 9 single-particle levels. In this case, we do not make the comparison with reaction NMEs.
 All NMEs are multiplied by $10^{3}$.
 }
\label{T4}
%\centering
\begin{small}
		\newcommand{\cc}[1]{\multicolumn{1}{c}{#1}}
		\renewcommand{\tabcolsep}{0.3pc} % enlarge column spacing
		\renewcommand{\arraystretch}{1.} % enlarge line spacing
\bigskip
\begin{tabular}{cccccccccccc}
\hline
\hline
& $0^+_f$& $M^{0\nu}_{V}$& $M^{0\nu}_{V_{0^+}}$& $\vert M _{00}\vert$& $M^{0\nu}_{A}$& $M^{0\nu}_{A_{1^+}}$  & $\vert M_{10}\vert$&  $M^{0\nu}_P$ &$M^{0\nu}_{M}$ &$M^{0\nu} $  &  $\langle m_{\nu} \rangle $  \\
 \hline
& $^{76}$Se          &          &           &          &          &                 &          &           &             &           &           \\
&                             &          &           &          &          &    P1         &        &           &            &           &               \\
& $0_1^+$ &       234&       145 &       221 &       946 &       466 &       399 &       -91 &        45 &      1133 & 0.69\\
& $0_2^+$ &       168 &       106 &       170 &       826 &       457 &       419 &       -72 &        33 &       955 & 60.9\\
   \hline
&         &           &           &          &          &    P2     &          &           &            &           &           \\
& $0_1^+$ &  -296 &    -198 &       305 &    -1176  &      -650 &     573&       105 &  -50&  -1417  &  0.55 \\
& $0_2^+$ &    211&      144 &       234 &      1030 &       635 &     594&       -82 &   36 &   1195  & 48.6 \\
 \hline
& $^{124}$Te          &          &           &          &          &          &          &           &             &           &           \\
&        &          &           &          &          &   7 levels          &          &           &            &           &           \\
%& $0_1^+$ &       34.1 &       &       &      133 &        &     &      - 13.7 &      8.0&      162 \\
& $0_1^+$ &       34 &       &       &      133 &        &     &      - 14&      8.0&      162 \\
& $0_2^+$ &        5.7&      &       &   6.8 &         &     &        -3.4 &   1.0&       10&      \\
%& $0_2^+$ &        5.71 &      &       &   6.78 &         &     &        -3.38 &   1.00&       10.1&      \\
   \hline
&            &          &           &          &          &   9 levels        &          &           &            &           &           \\
& $0_1^+$ &      42.9 &       &       &      142 &        &     &      - 15 &      8.4&      178 \\
& $0_2^+$ &     -9.3 &      &        &   -5.2 &         &     &        -3.0 &   -0.9&       -12&      \\
%& $0_2^+$ &     -9.32 &      &        &   -5.18 &         &     &        -3.0 &   -0.94&       -12.4&      \\
\hline\hline
\end{tabular}
\end{small}
\end{center}
\end{table}
%%%%%%

\subsection {\bf  DCE   Sum Rules, and Calibration of NMEs}
\label{Sec4D}

To verify the sum rules \rf{11} we have to evaluate the total reaction strengths $S^{\{\mp 2\}}_{J \J}$ from \rf{8} and \rf{16}, %, and \rf{50}
and their differences $ S^{\{2 \}}_{J \J}$ from \rf{11}. This will allow us to calibrate the NMEs 
$M_{J \J}$ from \rf{13}.  But, to calibrate
 in the same way  the 
 NMEs $M^{0\nu}_{V}$, and $M^{0\nu}_{A}$, we also need to know  the total and vector and axial vector strengths 
$ S^{0\nu^\mp}_{V,A} $, which are calculated from  \rf{32}, but with $ M^{0\nu^\mp}_V(0_f)$, and $M^{0\nu^\mp}_A(0_f)$ in place of $M^{0\nu^\mp}(0^+_f)$.

All results for  the DCE transition strengths in $^{76}$Ge, and   $^{124}$Xe are  summarized in Table \ref{T5}. We also show results  for the partial  vector and axial-vector strengths $ S^{0\nu^\mp}_{V_{0^+},{A_{1^+}}} $,
 which were defined in Section  \ref{Sec4C1}.

%\newpage
%%%%T4
\begin{table}[th]
\caption{Results for the DCER  
strengths $ S^{\{\mp2 \}}_{J\J}$, 
and  the $0\nu\b\b^\mp$-decay strengths $ S^{0\nu^\mp}_{V,A} $,
and $ S^{0\nu^\mp}_{V_{0^+},{A_{1^+}}} $, in  $^{76}$Ge, and   $^{124}$Xe. 
In  the case of $^{76}$Ge identical results are obtained with parameterizations 
P1 and P2. The inequalities in this table are due to the fact that the $C$  
terms in  \rf{11} are not included in the calculations.
}
\label{T5}
\begin{small}
		\newcommand{\cc}[1]{\multicolumn{1}{c}{#1}}
		\renewcommand{\tabcolsep}{0.35pc} % enlarge column spacing
		\renewcommand{\arraystretch}{1.} % enlarge line spacing
\begin{tabular}{cccccccccc}
\hline\hline
&{$J \J$} &{$ S^{\{{ -2}\}}_{J\J}$}& {$ S^{\{{ +2}\}}_{J \J}$ }&
{$ S^{\{2\}}_{J \J}$} &
{${\textsf S}^{\{2\}}_{J \J}$} &
%{${\bar E_{J}}^{\{-2\}}$}
$ S_{V,A}^{0\nu^-} $  &   $S_{V_{0^+},A_{1^+}}^{0\nu^-}$
&$ S_{V,A}^{0\nu^+} $  &   $S_{V_{0^+},A_{1^+}}^{0\nu^+}$\\\hline
&$^{76}$Ge&&&&&&&\\
%&&&&&&9 levels&&\\
&$00 $&$       300 $&$  0.56$&$       299 $&$       264  $&$       135 $&$       128 $&$  1.9 $&$  0.23 $\\
&$10 $&$       330 $&$  1.4 $&$       328 $&$   \le       353$&$       469 $&$       375 $&$  36$&$  1.6 $\\
&$12 $&$      1536 $&$  6.7$&$      1529 $&$   \ge   1403 $&$     $&$                 $&$       $&\\
\hline
%&&&&&$^{124}$Xe&7 levels&&\\
&$^{124}$Xe&&&&&&&\\
&&&&&&7 levels&&\\
&$00 $&$       549 $&$  1.5 $&$       548 $&$       480 $&$        221 $&$       217 $&$  2.3 $&$  0.55 $\\
&$10 $&$       641 $&$  6.7 $&$       634 $&$  \le      661 $&$        753 $&$       673 $&$  43 $&$  7.0 $\\
&$12 $&$      2996 $&$  31.0 $&$      2965 $&$ \ge      2827 $&$   $&$      $&$       $\\
\hline
%&&&&&$^{124}$Xe&9  levels&&\\
&&&&&&9  levels&&\\
&$00 $&$       558 $&$  2.0 $&$       556 $&$       480  $&$       231 $&$       220 $&$          3.3  $&$          0.71 $\\
&$10 $&$       653 $&$  8.0$&$       645 $&$   \le     672 $&$         826 $&$       685 $&$        57$&$            8.1 $\\
\hline\hline
\end{tabular}
\end{small}
\end{table}

We have  found that:

1) In  the case of $^{76}$Ge, identical results are obtained with parameterizations P1 and P2.

2) In the case of $^{124}$Xe, both calculations with 7 and 9 single-particle states satisfy the sum rules ~\rf{11}.

3) To satisfy the conditions in Eq. \rf{12}, it is essential to include the term $2S^{\{-1\}}_1$ in \rf{11};
otherwise, for example, in $^{76}$Ge, instead of ${\textsf S}^{\{2\}}_{10}=353$, 
we obtain ${\textsf S}^{\{2\}}_{10}=312$.

4)  Since ${\textsf S}^{\{2\}}_{J \J}$ are independent of the nuclear model, and  as the term proportional to $C$ is not taken into account when evaluating 
$ {\textsf S}^{\{ 2\}}_{10}$, and  
$ {\textsf S}^{\{ 2\}}_{12}$, the result:  

\hspace{1.5cm}$S^{\{2\}}_{00}\cong{\textsf S}^{\{2\}}_{00}$,
\hspace{.5cm}
$S^{\{2\}}_{10}\ge{\textsf S}^{\{2\}}_{10}$,
\hspace{0.5cm}
$S^{\{2\}}_{12}\le{\textsf S}^{\{2\}}_{12}$,

\noindent
implies that are satisfied both  the DCER  sum rules \rf{11}, and  the relations \rf{12}.
This,  in turn, means that the reaction transition strengths 
 $S^{\{-2\}}_{J \J}$, and $S^{\{+2\}}_{J \J}$ are  evaluated correctly. 
 Then, it is reasonable to assume that the results for the total transition 
 strengths $ S^{0\nu^\mp}_{V,A} $, and $ S^{0\nu^\mp}_{V_{0^+},{A_{1^+}}} $ 
 are also correct.

5) As the term proportional to $C$ is not taken into account when evaluating 
$ {\textsf S}^{\{ 2\}}_{10}$, 
and  $ {\textsf S}^{\{ 2\}}_{12}$ using \rf{11}, 
the inequalities \rf{12} are satisfied as expected. 
%\newpage

6) While the  partial strengths $ S^{0\nu^-}_{V_{0^+},{A_{1^+}}} $   are only slightly smaller than the total   strengths $S^{0\nu^-}_{V,A} $, 
 the  partial strengths               $ S^{0\nu^+}_{V_{0^+},{A_{1^+}}} $  are very small compared to  the total   strengths $ S^{0\nu^+}_{V,A} $.
In both cases the partial  strengths $ S^{0\nu^\mp}_{V_{0^+},{A_{1^+}}} $ are similar in magnitude to their reaction counterparts 
$ S^{\{{\mp 2}\}}_{J0}$. In particular, 
\brn
 { S}^{\{\mp 2\}}_{10}  &\cong& S^{0\nu^\mp}_{A_{1^+}},
%\label{45}
\ern
 which is consistent with $ \vert M_{10}\vert \cong \vert M^{0\nu}_{A_{1^+}}\vert$  reported earlier in Table \ref{T4} .
That is, the total transition strengths of the double Gamow-Teller NMEs are congruent to 
those of the $0\nu$ axial-vector NMEs that are engendered from 
the $1^+$ intermediate states (see also the Fig. \ref{F2}). 
 
7) From the above results we can now calibrate the NMEs, evaluating the amounts of the total strengths that are concentrated in the ground state, which are defined in Eqs. \rf{13} and \rf{33}.

They  are shown in Table \ref{T6} for $^{76}$Ge. All ratios $R$ are of the order 
of $10^{-4}-10^{-3}$. These results are
fully consistent with the findings of Auerbach \etal  \cite{Aue89,Aue18,Aue18a}, who concluded, 
from the shell-model  study of  the $ 2\nu\b\b^-$-decay in $^{48}$Ca, 
that for  the double Gamow-Teller is $ R^{-}_{10}(0^+_1) \sim (10^{-4}-10^{-3})$.

 The above ratios are even smaller in $ 0\nu ee$-capture processes.
For instance, for the ground state in $^{124}$Te with 7 levels 
we get $ R^{0\nu}=6.9 \cdot 10^{-4} $.
%%%%%%%%
\begin{table}[h]
\centering
\caption{\label{T6} Values of the ratios of the strengths defined in \rf{13} 
for the ground state in $^{76}$Se, using the parametrizations P1 and P2.
 All the ratios are multiplied by $10^{3}$.
}
                \begin{small}
		\newcommand{\cc}[1]{\multicolumn{1}{c}{#1}}
		\renewcommand{\tabcolsep}{1.7pc} % enlarge column spacing
		\renewcommand{\arraystretch}{1.} % enlarge line spacing
\begin{tabular}{cccccc}
\hline\hline
&$ R^-_{00}$&$  R^-_{10}$&$ R^{0\nu}_V $&$R^{0\nu}_A$    &$ R^{0\nu} $\\
\hline
%$^{76}$Se &     &        &&& \\
%  P1 & $2.92\cdot 10^{-4} $&$9.9\cdot 10^{-4} $&$6.4\cdot 10^{-3} $&$2.9\cdot 10^{-3} $& $2.1\cdot 10^{-3} $ \\
%  P2&$1.82\cdot 10^{-4} $&$6.6\cdot 10^{-4} $&$4.1\cdot 10^{-3} $ &$1.9\cdot 10^{-3} $&$3.3\cdot 10^{-3} $ \\
P1& $0.29 $&$0.99$&$6.4$&$2.9$& $2.1$ \\
P2& $0.18 $&$0.66$&$4.1$ &$1.9$&$3.3$ \\
\hline\hline
\end{tabular}
\end{small}
\end{table}
%%%%%
\begin{figure*}[]
\begin{center}
   \includegraphics[scale=0.95]{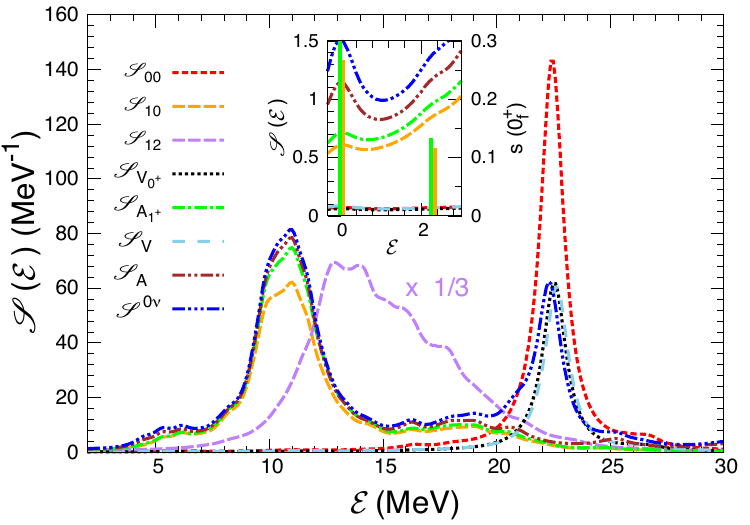}%v20Ge76-.eps}
\caption{ 
The folded strength distributions  ${\mathscr S}^{\{- 2\}}$ for the DCER  and  the $0\nu\b\b^-$-decay 
are shown as a function of the excitation energy $\E$ in the final nucleus $^ {76}$Se.   The intensities $ {\textsf s}^{\{{ -2}\}}_{10}$, $ {\textsf s}^{0\nu^-}_{A_{1^+}}$, and 
$ {\textsf s}^{0\nu^-}$  of the $0^+_{1,2}$ states are shown in the inset plot.
}
\label{F4}
\end{center}
\end{figure*}
%%%%%%
\subsection {\bf  DCE Giant Resonances }
\label{Sec4E}

The comparison made in Table \ref{T4} between NMEs is now extended
to the full strength distributions of the $0\nu\b\b$-decay and DCER NMEs, 
plotting their squares as a function of the excitation energies $\E_f$. 
Thus, we show in Figure~ \ref{F4} the folded transition densities ${\mathscr S}$, defined in \rf{34},
for the strengths  $ {\textsf s}_{00},  {\textsf s}_{10},  {\textsf s}_{12},  {\textsf s}^{0\nu^-}_{V_{0^+}},  
{\textsf s}^{0\nu^-}_{V}, {\textsf s}^{0\nu^-}_{A_{1^+}}, {\textsf s}^{0\nu^-}_{A}$, 
and $ {\textsf s}^{0\nu^-}$ in 
$^{76}$Ge, using the parametrization P2.

From this figure it is clear note that the  densities ${\mathscr S}^{\{- 2\}}_{00}$ 
and  ${\mathscr S}^{0\nu^-}_{V}\cong {\mathscr S}^{0\nu^-}_{V_{0^+}}$ behave 
similarly to each other, as do ${\mathscr S}^{\{- 2\}}_{10}$ 
and  ${\mathscr S}^{0\nu^-}_{A}\cong {\mathscr S}^{0\nu^-}_{A_{1^+}}$.
They are concentrated, respectively, at  $\sim   21$ MeV, and at $\sim 13$  MeV, 
where the corresponding DCE resonances, {\it i.e.}, the Double Isobaric Analog 
State (DIAS), and   the  Double Monopole Gamow-Teller 
Resonance (DMGTR), are localized. 

 Our DCETDA result for DIAS energy agrees well with the analytical estimate by Roca-Maza, Sagawa and Col\`o  \cite{Roc20} given by
\br
\E_{DIAS}=2\E_{IAS}+\Delta \E_{Coul},
\label{56}\er
where  $E_{IAS}$ is the energy of the isobaric analog state (IAS), defined as
\br
\E_{IAS}=\E_{Coul}(Z,A)-\E_{Coul}(Z,A-1).
\label{57}\er
The Coulomb energies are taken from  \cite[(2-19)]{Boh69}, \ie
\br
\E_{Coul}(Z,A)=0.70\frac{Z^2}{A^{1/3}} [1-0.76Z^{-2/3}] \, \mbox{MeV},
\label{58}\er
while   from \cite[(19)]{Roc20}
\br
\Delta \E_{Coul}\cong\frac{3}{2}A^{-1/3} \, \mbox{MeV}.
\label{59}\er
From Eq. \rf{56} we now obtain $\E_{DIAS}=2\x 10.21+0.35$ MeV $= 20.77$ MeV, which agrees perfectly with our nuclear model calculations,
%{\color{blue}  
given  in Fig.~ \ref{F4}   by the peak energy of ${\mathscr S}^{\{- 2\}}_{00}$  .

Also shown in the same figure the DCER quadrupole strength ${\mathscr S}^{\{- 2\}}_{12}$, and its double quadrupole Gamow-Teller resonance (DQGTR) at $ \sim 14$ MeV. As mentioned above, this resonance is not related to the Majorana mass. However, the comparison between calculations and experimental data are important to test the nuclear models.  

The total density distribution ${\mathscr S}^{0\nu^-}$ is the envelope 
of ${\mathscr S}^{0\nu^-}_{V}$,  and  ${\mathscr S}^{0\nu^-}_{A}$,  
including also the pseudoscalar and weak magnetism contributions.
Very similar spectra are obtained when the parameterization P1 is used, 
with the main differences being: (i)  
the axial vector strength now peaks at $\sim 15$ MeV, and (ii) the insertion 
plot is modified somewhat, as indicated in Table \ref{T4}.

Needless to say, only the NMEs that are within the $Q$-value window 
(shown in the inset of Figure \ref{F3}), are significant to the neutrino mass. 
It is relevant to note in this insert the similarity of $ {\textsf s}^{\{{ -2}\}}_{10}$ 
with $ {\textsf s}^{0\nu^-}_{A_{1^+}}$ and its difference with  
$ {\textsf s}^{0\nu^-}$. Of course, the same issue can be seen in Table \ref{T3}.

However, the entire 
reaction spectrum $({{\mathscr S}}^{\{-2\}}_{00}+{{\mathscr S}}^{\{-2\}}_{10})$  
can, in principle, be measured, and especially in the region of  DIAS,  
and DMGTR resonances, which are well separated from each other. It would then be 
worthwhile to perform such a measurement, and compare the resulting data with 
theoretical predictions, in line with what was done in Ref. ~\cite{Yak09} for the two GT 
spectra ${\mathscr S}^{\{\mp 1\}}_{10}$ in $^{48}$Sc, and in  Ref. \cite{Yas18} for the  GT 
spectrum ${\mathscr S}^{\{- 1\}}_{10}$ in $^{132}$Sb.
Agreement between such measurements with the calculations shown in Fig. \ref{F4}
them would test our nuclear structure model,
as well as that the predictions about the NMEs make sense. 

\subsection{\bf Relationships between calculations and reaction data}
\label{Sec4F}

The question posed here is whether the DCE transition strengths ${{\mathscr S}}^{\{\pm 2\}}_{J\J}$ are experimentally accessible, in the same way as the SCE strengths ${\mathscr S}^{\{\pm 1\}}_{10}$~\cite{Yak09,Yas18} are ?

It should be recalled that in the SCE case it is possible to study GT transitions via light-ion–induced reactions, such as (n,p), (p,n), ($^3$He,t), and (t,$^3$He), thanks to the proportionality relationship between the zero angular momentum transfer cross section, $\sigma$ (at a forward angle) and the corresponding GT strength $ {\textsf s}^{\{- 1\}}_{1}$ (see  \cite[Eq. (3)]{Yak09}, and   \cite[Eq. (1)]{Yas18}). Namely,
\br
\sigma_{SCE-GT}\sim  {\mathscr S}^{\{ -1\}}_{1},
\label{60}\er
where the strength density $ {\mathscr S}^{\{- 1\}}_{1}$ in the odd-odd nucleus is defined in the same way as 
${\mathscr S_{J\J}^{\{- 2\}}}$ in Eq.
\rf{38} 
for the final nucleus, with ${{\textsf s} ^{\{\mp 1\}}_{J}(J_i)}$
instead of ${{\textsf s} ^{\{\mp 2\}}_{J\J}(\J_f)}$.

In the DCE case happens something similar in the $( \pi^+, \pi^-)$ reactions, but only with Fermi transitions, for which 
\br
\sigma_{DCE-F}\sim  {\textsf S}^{\{ 2\}}_{00}\cong S^{\{- 2\}}_{00},
\label{61}\er
as demonstrated in \cite[Fig. 1]{Kal87} (see also \cite{Aue87}).
 Moreover, in the spectrum $
{\mathscr S}_{00}(\E)$, the DIAS resonance is clearly perceived at the energy $\E_{DIAS}$, given by Eq. \rf{56}, as can be seen in \cite[Fig. 2]{Aue90}.
Moreover, the  Fig. \ref{F4} shows that the DIAS is a rather narrow resonance in the Fermi spectra ${\mathscr S}_{00}(\E)$, suggesting that near its maximum at $\E_{DIAS}$, it behaves like 
\br
\sigma_{DCE-F}\sim {\mathscr S}^{\{- 2\}}_{00}(\E).
\label{62}\er
The pion, however, interacts weakly with states involving the spin degree of freedom and therefore states such as   DMGTR and  DQGTR
 are not observed in $( \pi^+, \pi^-)$ reactions.

As mentioned above, heavy ion DCE reactions are being carried out using light ions such as $^{18}$O.  It is hoped to learn something about the magnitude of the  $0\nu\b\b$ NMEs by measuring the  $M_{00}$, and $M_{10}$, whose weak analogs  $M^{0\nu}_{V_{0^+}}$, and  $M^{0\nu}_{A_{1^+}}$ are the major contributor to them, as seen in Table \ref{T4}, and discussed in detail above.

The difficulty is due these heavy ion DCE reactions are much more complicated than those of $( \pi^+, \pi^-)$, especially since they probe the NMEs of the projectile and the target at the same time, requiring additional theoretical efforts to disentangle the two types of contributions \cite{Len19,Bel20,Cap15}. In this scenario, the Gawow-Teller cross section cannot be expressed in the form of the Fermi cross section, given by Eq. \rf{61}.
%\newpage

 Based on a previous work of Bertulani~\cite{Ber93},  Santopinto \etal ~\cite{San18} have recently reported that,  for the ground-state target nuclei
%Recently, Santopinto \etal \cite{San18} have reported that,  for the ground-state target nuclei
  the DCE differential cross section within the low-momentum-transfer limit can be factored  into:  i) a reaction part, which  is computed by means of the eikonal approximation, and ii) the nuclear part
\br
 {\mathscr N}=\left\vert
                                                         \dfrac{M_{00}^PM^T_{00}}
{{\bar E^{P}}_{0}+{\bar E^{T}}_{0}}
                                                         +\dfrac{M_{10}^PM^T_{10}}
{{\bar E^{P}}_{1}+{\bar E^{T}}_{1}}\right \vert^2,
\label{63}\er
where $P$ and $T$ stand for projectile and target nuclei respectively, and
\br
 \bar{E}_{J} =\dfrac{ \sum_{i}
{ {\textsf s}^{\{-1\}}_{J}(J_i)}(E_{J_i }^{\{- 1\}}-E_{0^+}^{\{0\}})}{S^{\{- 1\}}_{J}}
\label{64}\er
 are the averaged energies of the intermediate nucleus (see \rf{1}).  The purpose is to put an upper limit on $M^T_{10}$, which will correspond to an upper limit on $M^{0\nu}$ for  $\b\b$-decaying nuclei. 
Something similar has been done previously by  Cappuzzello \etal ~\cite{Cap15} for the  reaction $^{40}$Ca($^{18}$O, $^{18}$Ne)$^{40}$Ar, 
in which the $\b\b$-decay does not occur.

All NMEs and energies in \rf{63} can be  calculated straightforward in the DCEQTDA,  and we only need the experimental values of $ {\mathscr N}$ to make the comparison with the data. This would be the first step  toward the calibration of  $M^{0\nu}$ by heavy ion reaction experiments.

Finally, it should be noted that only at low momentum transfer are the NMEs in heavy-ion reactions of the form of Eq. \rf{16}, where only the intermediate $J^\pi =0^+,1^+$ states contribute, as in $ 2\nu\b\b$-decay. 
 In fact, as pointed out in Ref.  \cite{Bel20}, the  transition amplitude requires in 
 %{\color{red}  this} {\color{blue}  
 the general
 %} 
case the inclusion of all multipole terms with $J^\pi=0^\pm,1^\pm, \cdots 10^\pm$, since the momentum transferred  is of the order of $q\geq 400$  MeV/c as in the $ 0\nu\b\b$ decay, where $q\sim 100$  MeV/c  \cite{Tom91}.

\section{ Summary and Final Remarks}
\label{Sec5}
In Section 1 we briefly state that our main goal is to assist the detection of the Majorana neutrino by observing $0\nu\b\b$-decay,
which will allow us to find the effective mass of the neutrino $\langle m_ {\nu}\rangle$, provided we know the nuclear matrix element  
$M^{0\nu}$.

In Section 2 we present the general formalism for the theoretical description of DCE processes described by two-body DCE operators.

In Section 3 we complete the formulation of the DCEQTDA, which was proposed some time ago \cite{Krm05}, and developed in detail for application to the $2\nu\b\b$-decay \cite{Fer17}. Here,  we extend it to the $0\nu\b\b$-decay.

In Section 4 we use the DCEQTDA formalism to describe different DCE observables, with special emphasis on the  $^{76}$Ge, and 
  $^{124}$Xe.
  First, in Section \ref{Sec4B} we study all the DCE observables that have been measured so far, both the static ones (excitation energies in the final nuclei, and the $Q_{\b\b^-}$ and $Q_{ee}$ values), and the dynamic ones which are the $\tau_{2\nu}$ half-lives of all $0^+$ and $2^+$ states that lie within the windows of $Q$-values, obtaining good agreement with the experimental data for all of them.

After  ensuring in this way the reasonableness of the DCEQTDA, we use it  in Section \ref{Sec4C} to study  the  $0\nu\b\b$-decays and   
$0\nu ee$-captures in a series of nuclei.  We start in Section \ref{Sec4C1} with the
analysis of the genesis of the  $0\nu\b\b$-decay  going to the ground state in $^{76}$Se, comparing it with the results of the $pn$-QRPA. 
We find that these two nuclear models are physically very different and that the main reason for this is the approximation 
 $\Bra{\J^+_f}\O^{\pm}_J\Ket{J^{+}_i}\cong
\Bra{\bar J^{+}_i}\O^\mp_J\Ket{\bar 0^{+}}$ performed in the $pn$-QRPA, as discussed in Section \ref{Sec3}.

 In section \ref{Sec4C2}  we make a detailed comparison between the $pn$-QRPA and DCEQTDA calculations for the $0\nu\b\b$, and  $0\nu ee$ NMEs of the final ground state. In the same figure, we also show the results of several other theoretical studies, observing that in DCEQTDA the matrix elements turn out to be significantly smaller than in other nuclear models, with the exception of Ab initio models.

In Section \ref{Sec4C3} we point out that it is not correct to associate the reaction NMEs $M _{00}$ and  $M_{10}$ with the $0\nu\b\b$ NMEs $M^{0\nu}_{V}$, and $M^{0\nu}_{A}$,  as is often done.

In Section \ref{Sec4D} we have calculated total strengths $ S^{\{\mp2 \}}_{J\J}$  in $^{76}$Ge, and   $^{124}$Xe. So far, this has been done only for light nuclei up to $^{48}$Ca \cite{Aue18,Shi18}.
To calibrate the $0\nu\b\b$  NMEs we have also done so for the total strengths $ S^{0\nu^\mp}_{V,A} $, which has never been done before.
We conclude that the fractions of $ S^{\{\mp 2 \}}_{J\J}$ and $ S^{0\nu^\mp}_{V,A} $  are of the order of  $10^{-4}-10^{-3}$ in $0\nu\b\b$-decays and still smaller in $ 0\nu ee$-capture processes.

The comparison made in Table \ref{T4} between reaction and $0\nu\b\b$ NMEs is extended in Section \ref{Sec4E}
 by exposing in Fig. \ref{F4} the full spectral densities ${\mathscr S}^{\{- 2\}}_{J\J}$,   and  ${\mathscr S}^{0\nu^-}_{V,A}$, which also include the giant DCE resonances.
The fact that they turned out to be of the same order came as a surprise, as we have found no physical reason to justify it.

We have advanced to the energy region where giant DCE resonances should be found, aware that $\b\b$-decays cannot occur there. However, experimental verification of our prediction would reinforce confidence in the theoretical comparison made.

Although the main motive of this work has been the study of $\b\b$-decays, and in particular of the  $0\nu$-mode, in Section \ref{Sec4F} we also briefly discuss nuclear reaction experiments, which are related to this decay mode.

 We demonstrate that theoretical methods are at hand, ready to describe the DCE reaction data at low momentum transfer that will be available in the near future. 

The present nuclear structure model contains the same free parameters as the $pn$-QRPA model and is specially designed to describe double charge exchange processes. As such it has the following features:

 1) It calculates the wave functions in the final nuclei, which contains four-quasiparticle excitations and the Pauli Principle, among them,  which play a crucial role,

2) It allows working with a single-particle space large enough for the DCE sum rules to be satisfied; and 

3) it jointly describes the $\b\b$-decays and DCERs to all $0^+$ and $2^+$ final states, 
as well as their $Q$-values, energies, resonances, and sum rules. 

We believe that these three aspects are indispensable for a reliable assessment 
of  $ 0\nu\b\b$  NMEs. The commonly used $pn$-QRPA model, being limited to the description 
of ground-state NMEs, does not meet these conditions.  
As far as we know, the only other nuclear model capable of evaluating all the double charge exchange observables, which we discuss here, is the Shell Model, but such a comprehensive study has not been done. So far, only Auerbach and his collaborators \cite{Zhe89,Aue89,Aue18,Aue18a, Aue87,Aue90} have gone in this direction.	

%A more complete calculation  of these  NMEs will be carried out by analyzing all nuclei 
%in which double beta and double electron capture processes occur.

%\begin{acknowledgements}
\section*{Acknowledgments}
%We sincerely thank Wayne Seale and Tom T. S. Kuo for their careful and enlightening reading of the manuscript.
We sincerely thank Wayne Seale for the careful and enlightening reading of the manuscript.
This work was financed in part by the Coordena\c{c}\~ao de Aperfei\c{c}oamento
de Pessoal de N\'ivel Superior Brasil %(CAPES)
Finance Code 001.
A.R.S. acknowledges the financial support of 
UESC (SEI 073.6766.2020.0010299-61)
and Fundação de Amparo \`a Pesquisa
do Estado da Bahia
TO PIE013/2016. V. dos S. F.
acknowledges the financial support of UESC-PROBOL program.
%\end{acknowledgements}

\appendix
%%%%%%%%%%%%%%%%%%%%
\section{Appendix}\label{}

\begin{table}[th]
\caption{Comparison of the DCEQTDA calculations to data for the $2\nu\b\b$ NMEs, and the DCE Q-values $Q_{\b\b}$, and $Q_{ee}$
in nuclei:  $^{48}$Ca, $^{82}$Se,  $^{96}$Zr,   $^{100}$Mo, $^{116}$Cd, $^{128}$Te, $^{130}$Te, $^{146}$Xe,  $^{150}$Nd, $^{78}$Kr , and   $^{130}$Ba.   The experimental values of $M^{2\nu}$  correspond to the $G$ factors  from Ref. \cite{Pir15}.  }
 
\label{T7}
\begin{tabular}{ cc cc cc cc }
\hline\hline
Nucleus   &$\J_f^+$   &&$ M^{2\nu}   $&&$Q_{\b\b}$&&$Q_{ee}$\\%\hline
&& cal&exp  & cal&exp & cal&exp \\\hline
$^{48}$Ca &$  0^+_1                   $&$    0.107  $&$   0.038$&$   4.016        $& $ 4.268$  &$-17.383 $  &$  -21.900 $ \\%\hline
%$^{48}$Ca    &$ 0^+_1      $&$    0.107  $&$   0.038  $&$   4.016   $&$ 4.268  $ &$  -17.383  $&$  -21.900   $\\\hline
 %$^{76}$Ge &$  0^+_1                   $&$    0.103  $&$   0.107$&$   1.314       $& $ 2.039$  &$-8.165 $&$  -10.909 $ \\
 %                  &$0^+_2                      $&$    0.158  $&$  0.106  $&$        --      $& $      -    $ &$   --  $&$    -- $\\\hline
$^{82}$Se  &$  0^+_1                   $&$    0.031 $&$  0.085  $&$   2.155  $&$    2.997 $&$   -10.459 $&$   -12.178  $\\%\hline
$^{96}$Zr&$  0^+_1                   $&$    0.134 $&$  0.080 $&$   2.537  $&$    3.356 $&$   -8.939  $&$ -12.516$\\%\hline
%       0   1     0.351E-02   0.130E+00   0.134E+00     0.080     2.537     3.356    -8.939   -12.516
   $^{100}$Mo &$  0^+_1               $&$    0.084  $&$   0.210$&$ 0.918 $&$        3.034 $&$       -9.117 $&$       -9.816$ \\
&       $0^+_2                      $&$    0.599  $&$  0.151  $&$        -      $& $      -    $ &$   -  $&$    - $\\%\hline
$^{116}$Cd &$  0^+_1               $&$    0.030  $&$   0.108$&$  1.322 $&$     2.813 $&$    -8.969 $&$    -8.882$ \\%\hline
$^{128}$Te &$  0^+_1                 $&$    0.040 $&$  0.045$&$  -1.101     $&$    2.813  $&$   -7.107 $&$    -8.881 $\\%\hline
%$^{128}$Te &$  0^+_1                   $&$    0.043 $&$  0.046  $&$   -1.162  $&$     0.867 $&$     -4.928  $&$    -5.632 $\\%\hline
$^{130}$Te &$  0^+_1                   $&$    0.040 $&$   0.030  $&$   0.447  $&$    2.527  $&$   -6.389  $&$   -7.221$\\%\hline
   %$%^{116}$Cd &$  0^+_1                   $&$    0.040 $&$  0.116   $&$  -1.101 $ &$     2.813  $ &$   -7.107   $ &$  -8.881 $\\\hline
 $^{146}$Xe &$  0^+_1               $&$  0.024   $&$  0.018  $&$ 1.147  $&$ 2.457   $&$  -9.555    $&$ -12.003$ \\%\hline
 $^{150}$Nd &$  0^+_1               $&$          0.041  $&$    0.045 $&$     3.406  $&$     3.371  $&$    -9.290  $&$    -8.833$ \\
 &       $0^+_2                      $&$  0.033  $&$  0.071  $&$        -      $& $      -    $ &$   -  $&$    - $\\%\hline
$^{78}$Kr &$  0^+_1   $&$    0.031  $&$   0.358$&$   -13.484  $&$  -11.004  $&$    5.181   $&$   2.848$ \\%\hline
  $^{130}$Ba &$  0^+_1 $&$    0.039 $&$     0.175 $&$    -9.399  $&$   -7.840 $&$  4.726  $&$    2.619$ \\\hline\hline
\end{tabular}
\end{table}

 Table   \ref{T7} compares the DCEQTDA calculations with the experimental data for the  $2\nu\b\b$ NMEs, and for the DCE Q-values $Q_{\b}$, and $Q_{ee}$, in nuclei other than $^{76}$Ge, and $^{124}$Xe. 
In the calculation we have used 9 single-particle levels in all nuclei except $^{48}$Ca, where 7 levels were used. The singlet and triplet coupling constants of the delta force are those of parameterization P1,   \ie $s=s_{sym}=1$, and $t=t_{sym}   \gtrapprox 1$.

As indicated in  Eq. \rf{31} the experimental NMEs $M^{2\nu}  $  depends on both the measured half-life
$\tau^{2\nu}$, and the calculated kinematical factor $G^{2\nu}$.
% {\color{red} Thus,} 
 The values of $M^{2\nu}(exp)$ listed in Table \ref{T7}  correspond to the $G$ factors  from Ref.\cite{Pir15}.   Note that our $M^{2\nu}(exp)  $  is in agreement with the $M^{eff}  $  of  \cite{Bar20}.
 
From the last  table it is clear that the agreement between the calculations and the experimental data is quite satisfactory for both NMEs and $Q$-values, especially if one takes into account that the parameters of the nuclear model were not varied. Indeed, except for  $^{100}$Mo, $^{78}$Kr , and   $^{130}$Ba, the calculated and measured NMEs are quite close to each other. It is also important that the values of $Q_{\b\b}$, both calculated and experimental, are both positive in nuclei decaying by $\b\b$, except at $^{128}$Te, and negative in nuclei where $ee$ capture occurs. The opposite is true in all cases with $Q_{ee}$.

%%%%%%%%%%
%END MAIN TEXT
%%%%%%%%%%

%newbibliography-Arturo31082024
%\section*{References}
%\bibliography{ref3}

%
\end{document}